\documentclass[traditabstract]{aa} % for the abstract without structuration
\usepackage{graphicx}
\def\gtrsim{\mathrel{\hbox{\rlap{\hbox{\lower4pt\hbox{$\sim$}}}\hbox{$>$}}}}
\def\ltsim{\mathrel{\hbox{\rlap{\hbox{\lower4pt\hbox{$\sim$}}}\hbox{$<$}}}}

\begin{document}
   \title{The relationship between
$\gamma$\,Cassiopeiae's X-ray emission and its circumstellar environment}

%   \subtitle{I. Overviewing the $\kappa$-mechanism}

   \author{
   M. A. Smith\inst{1}
        \and
   R. Lopes de Oliveira\inst{2,3}
        \and
   C. Motch\inst{4}
        \and
   G. W. Henry\inst{5}
        \and
   N. D. Richardson\inst{6}
        \and
   K. S. Bjorkman\inst{7}
        \and
   Ph. Stee\inst{8}, D. Mourard\inst{8}
        \and
   J. D. Monnier\inst{9}, X. Che\inst{9}
        \and
   R. B\"ucke\inst{10}
        \and
   E. Pollmann\inst{11}
        \and
   D. R. Gies\inst{6}, G. H., Schaefer\inst{6}, T. ten Brummelaar\inst{6},
   H. A. McAlister\inst{6}, N. H.  Turner\inst{6}, J. Sturmann\inst{6},
   L. Sturmann\inst{6}
        \and
   S. T. Ridgway\inst{12}
}

   \institute{Catholic University of America,
3700 San Martin Dr.,
Baltimore, MD 21218  USA; \email{msmith@stsci.edu,}  % [1]
        \and
             Universidade Federal de Sergipe, Departamento de F\'isica, Av. Marechal Rondon s/n, 49100-000 S\~ao Crist\'ov\~ao, SE, Brazil  % [2]
%             \email{...}
  \and
Universidade de S\~ao Paulo, Instituto de F\'isica de S\~ao Carlos, Caixa Postal 369, 13560-970, S\~ao Carlos, SP, Brazil %[3]
\and 
Universit\'e de Strasbourg, CNRS, Observatoire Astronomique,
11 rue de l'Universit\'e, 6000 Strasbourg, France  % [4]
  \and
Center of Excellence in Information Systems, Tennessee State
University, 3500 John  Merritt Blvd., Nashville, TN % 37203 % [5]
  \and  
Center for High Angular Resolution Astronomy, Department of Physics
and Astronomy, Georgia State University, P.O. Box 4106, Atlanta, GA
30202-4106, USA  %  [6]
  \and
Ritter Astrophysical Research Center, Department of Physics \&
Astronomy, University of Toledo, 2801 W. Bancroft, Toledo, OH 43606, USA % [7]
  \and
Laboratoire Lagrange, UMR 7293 UNS-CNRS-OCA, Blvd. l'Observatorie, B. P.  4229F,
06304 Nice Cedex 4, France % [8]
%UMR 6525 CNRS H. FIZEAU  UNS, OCA, CNRS, Campus Valrose, F-06108 Nice cedex 2,
%France  % [8]
  \and
 Department of Astronomy, University of Michigan, Ann Arbor, MI 48109, USA % [9]
  \and
  Anna-von-Gierke-Ring 147, 21035 Hamburg, Germany  % [10]
  \and
Emil-Nolde-Str. 12, 51375 Leverkusen, Germany   % [11]
  \and
National Optical Astronomical Observatory, 950 North Cherry Ave., Tucson, AZ
85719, USA  %  [12]
             }

   \date{Received October dd, 2011; accepted ????? DD, YYYY}

\authorrunning{Smith et al.}  \titlerunning{$\gamma$\,Cassiopeiae's properties}
  \abstract
{$\gamma$\,Cas is the prototypical classical Be star and is recently best known
for its variable hard X-ray emission. To elucidate the reasons for this 
emission, we mounted a multiwavelength campaign in 2010 centered around four 
{\it XMM-Newton}
observations.  The observational techniques included long baseline optical 
interferometry (LBOI) from two instruments at CHARA, photometry carried out 
by an Automated Photometric Telescope and H$\alpha$ 
observations. Because $\gamma$\,Cas is also known to be in a binary, we
measured radial velocities from the H$\alpha$ line and redetermined its 
period as 203.55${\pm 0.20}$ days and its 
eccentricity as near zero.  The LBOI observations suggest that the star's 
decretion disk was axisymmetric in 2010, has an system inclination angle near 
45$^{\circ}$, and a larger radius than previously reported. In addition,
the Be star began an ``outburst" at the beginning of our campaign, made visible 
by a brightening and reddening of the disk during our campaign and beyond.
 Our analyses of the new high resolution spectra disclosed many attributes 
also found from spectra obtained in 2001 (Chandra) and 2004 (XMM-Newton). 
As well as a dominant
hot ($\approx$ 14\,keV) thermal component, the familiar ones included:
(i) a fluorescent feature of Fe K even stronger than observed at previous times,
(ii) strong lines of N\,VII and Ne\,XI lines indicative of overabundances, and 
(iii) a subsolar Fe abundance from K-shell lines but a solar abundance from 
L-shell ions. We also found that two absorption
columns are required to fit the continuum. While the first one maintained its
historical average of 1$\times$10$^{21}$ cm$^{-2}$, the second was very large 
and doubled to 7.4$\times$10$^{23}$ cm$^{-2}$ during our X-ray observations.
Although we found no clear relation between this column density 
and orbital phase, it correlates well with the 
disk brightening and reddening both in the 2010 and earlier observations. Thus,
the inference from this study is that 
much (perhaps all?) of the X-ray emission from this source originates behind 
matter ejected by $\gamma$\,Cas into our line of sight.
}

%Stars: winds, outflows }

   \maketitle
%________________________________________________________________

\section{Introduction}
\label{intro}

   Discovered in 1866 by Secchi from the emission in its
H$\alpha$ spectral line, $\gamma$\,Cas became the first member of the
now well known group of ``classical Be" variables (Porter \& Rivinius
2003). These stars are also known for their rapid rotation, continuum 
excess in the infrared, and by the presence of strong, blueshifted, and 
often variable, absorption components in their UV resonance lines.
The latter were studied extensively in $\gamma$\,Cas and other Be stars 
at times intensively during the 1978-1996 operation of the
{\it International Ultraviolet Explorer (IUE).} Fortunately, soon 
after of the termination of the operations of this satellite, other 
techniques were being introduced to astronomy and have advanced the
study of the Be stars and their ejected Keplerian disks. These developments 
include the operation of Long Baseline Optical Interferometry (LBOI)
facilities, the development of automated photometric 
(robotic) telescopes, and an increasingly relevant contribution in
spectroscopy by amateur telescopes with their own local facilities.
The present study makes use of these elements.

Through the last decade astronomers have mounted a series of 
multi-wavelength campaigns on the prototype of the ``classical" Be 
stars, $\gamma$\,Cas (HD\,5394, type B0.5IVe). Even more than most
classical Be stars, $\gamma$\,Cas is variable over a number of wavelengths 
and timescales -- see Harmanec (2002).
The study of $\gamma$\,Cas developed renewed
focus by the discovery that it emits X-rays with unusual properties 
and in particular by White et al.'s (1982) suggestion that these
emissions arise from infall onto a degenerate companion. 
We now know that this star is in fact the primary in a nearly 
circular, 203-day binary system (Harmanec et al. 2000, Miroshnichenko 
et al. 2002). 
It may be important that its spectral lines are highly rotationally
broadened (vsin\,$i$ $\approx$ 400--441 km\,s$^{-1}$; Harmanec 2002, Fre\'mat
et al. 2005). 
With this broadening and a probable rotational period of 1.2158 days (Smith, 
Henry, \& Vishniac 2006; ``SHV"), this star is rotating at nearly its 
critical rate.

 $\gamma$\,Cas is also the prototype of a group that so far includes eight
other Galactic ``$\gamma$\,Cas X-ray analogs." These include:
HD\,110432 (Smith \& Balona 2006; ``SB06"), SS\,397, ``Star 9" in NGC\,6649
(USNO 0750-13549725; Motch et al. 2007), HD\,161103
\& SAO\,49725 (Lopes de Oliveira et al. 2006; ``L06"),
HD\,119682 (Rakowski et al. 2006, Safi-Harb et al. 2006), 
XGPS-36 (Motch et al. 2010), and
HD\,157832 (Lopes de Oliveira \& Motch 2011).  The X-ray properties
of this group are not shared by the other classical Be stars.

   In the years since the White et al. study, much effort
has been devoted to characterizing the X-ray attributes of $\gamma$\,Cas
itself because of its brightness. We know now that these 
properties include: (1) a rather high mean L$_x$/L 
$\sim$ 10$^{-6}$ (but at least 10 times lower than Be X-ray binaries),
(2) a light curve variable on a number of timescales, 
and (3) a thermal
spectrum consisting of 3-4 distinct plasma components. It is dominated by a 
hot component with temperature k$T_{hot}$ of 12--14 keV and subject 
to absorption by two independent columns of cold (ambient) gas
(Smith et al. 2004, Lopes de Oliveira 2010). 
It should also be remarked that the X-ray emission consists of an 
underlying ``basal" component, which varies on a timescale of hours, 
punctuated by rapid flares that according to plasma cooling arguments
alone must be produced in a high density environment (Smith, Robinson, \& 
Corbet 1998, ``SRC").\footnote{We define a flare as a sudden and 
short-lived increase in X-ray flux. This term does not necessarily 
connote a magnetic origin for them as on the Sun.}

The X-ray light curve also exhibits variations on at least three timescales. 
Much of the behavior of the X-ray light curve of $\gamma$\,Cas is now
well investigated by a series of relatively homogeneous {\it RXTE} 
studies (SRC, Smith \& Robinson 1999, Robinson \& Smith 2000, Robinson,
Smith, \& Henry (2006; ``RSH"), and SHV).
A rapid variability, first described in detail by 
Murakami et al. (1986), takes the form of ubiquitous flares on 
timescales of 2--3 minutes down to 4 seconds (the instrumental
detection threshold).  These flares are ``mild," only seldom having 
amplitudes of more than 2$\times$ above the background ``basal" flux.
The basal fluxes undergo undulations on timescales as short as 20 minutes
but more typically a few hours.
Thirdly, a long X-ray cycle of $\sim$70 day observed by RSH and SHV
seem to be well correlated with long periods at optical wavelengths. The
total range of these cycles in the optical range is about 50--91 days.

 In this paper we seek to understand the X-ray emission generation 
in terms of its possible relationship to the general Be circumstellar 
environment, including both the decretion disk and its widely spaced binary 
companion. The impetus for this study was the different amounts of soft 
X-ray flux in earlier high-resolution spectra by {\it Chandra} and the 
{\it XMM-Newton} satellites  (Smith et al. 2004, Lopes de Oliveira et al. 
2010).  The different soft X-ray flux levels are due to a density range
(by a factor of 40) in one of two absorption columns used in our models 
to fit the full {\it XMM-Newton} spectrum (the other column length 
solution was stable).  We noticed that the 2001 {\it Chandra}
observations were obtained during the binary phase of inferior 
conjunction, that is with the Be star passing through the foreground 
in its orbit. To see if the column density depends on binary phase,
we requested four new exposures in 2010 for inferior conjunction phase
and extending almost to quadrature. The 
question we posed was whether extra absorption would recur at 
this phase, e.g. due to absorption by matter near the Be star.

\section{Description of the circumstellar matter}
\label{dscrp}

Many of the optical and X-ray variations from 
$\gamma$\,Cas are likely to be caused somehow by interactions of 
matter released by the star with its disk or with previous ejecta. 
Since the present study sheds new light on some these variations
we should first review the status of the circumstellar (CS) structures. 
\\

\noindent {\it Geometry of the disk:~} \\
  For several decades the 
emission from the decretion disk of $\gamma$\,Cas has varied irregularly, 
presumably because of discontinuous feeding of matter to its decretion disk 
(Doazan 1982, Doazan et al. 1987). 
The emission in the continuum is formed by free-free and bound-free 
process and in the hydrogen lines by recombination. Stee \& Bittar (2001)
found that the disk is optically thin at visible wavelengths but optically 
thick in the near infrared and in the H$\alpha$ line out to greater disk radii. 
Their results indicated that the disk contributes 
$\gtrsim$ 10\% to the total flux of the star-disk system in the Johnson $V$ 
band but only 3--4\% in $B$. 
Spectroscopic and interferometric arguments suggest that the disk and 
rotational axes coincide, and both techniques 
indicate that this angle, the inclination, has an intermediate value. 
For example, in their initial interferometric work in H$\alpha$, Mourard
et al. (1989) determined a disk inclination angle of $\sim$45$^{\circ}$ from 
its measured ratio of the minor to major axis on the sky. Observations at this 
wavelength by Quirrenbach et al. (1997) and Tycner et al. 2006), with the
MkIII and NPOI interferometers, respectively, as well as in
the 2.1\,\,$\mu$m K' continuum band by Gies et al. (2007) have determined
inclinations in a range $i$ = 46--55$^o$. 

 LBOI studies have also determined the disk size of $\gamma$\,Cas. 
Tycner et al. H$\alpha$ found a diameter of the major axis of 3.67\,mas. 
{\bf This compared well
with the Quirrenbach et al. (1997) measurement of 3.47\,mas. 
In this paper we will adopt a uniform angular diameter for the
Be star of 0.44\,mas. Therefore, 
these measurements correspond to about 8\,R$_*$.}
%Assuming a distance of 
%{\bf 168\,pc (van Leuuwen (2007)}
%\footnote{We are aware of the revised distance of 
%168\,pc for $\gamma$\,Cas impled by the revised Hipparcos scale by van 
%Leeuwen (2007). For consistency with past measures we refer to the former 
%value. The revised value increases the measured linear sizes of the Be disk 
%we quote but is otherwise not significant to this study.}
%and a stellar radius of 10\,R$_{*}$ {\bf for \gamma$\,Cas,} we find a disk 
%radius of 6.6\,R$_*$. 
In addition, ost kinematic studies of Be disks have found that they
have Keplerian rotation (e.g., Meilland et al. 2007). Mourard et al. 
(1989) could show that the $\gamma$\,Cas disk rotates but were unable
to specify its velocity-distance relation.
\\

\noindent {\it Nonsecular variations imposed by the disk:~}  \\
 Historically $\gamma$\,Cas has been
among the ${\frac 14}$ of classical Be stars for which the violet (``$V$") 
and red (``$R$") components of their Balmer emission lines exhibit cyclic 
variations with timescales of years (e.g., Hummel 2000). These cycles are
believed to be caused by the precession of a 1-armed density structure from
the excitation of nonaxisymmetric disk modes (Okazaki 1991).
They were first observed in Balmer lines of $\gamma$\,Cas in 1969  
(Cowley, Rogers, \& Hutchings 1976).  Continued observations in the 
1980s--1990s disclosed a strong cycle with a length of 5--6 years. 
Berio et al. (1999) traced the precession of this arm over 
a few years by LBOI observations in H$\alpha$.
Ritter Observatory spectra obtained by KSB show that the 
cycle was still present in 2000. Later observations by Rivinius (2007) 
indicate that the cycle weakened through 2003. By 2005 the $V/R$ variations 
recorded by Ondrejov Observatory spectra were ``very weak if present at all" 
(Nemravov\'a 2011). Spectra obtained through the Be Star database (''BeSS",
Neiner et al. 2011) from 2009 to the present, 
including the epoch of our X-ray observations, show no variations either. 

  Typical of other classical Be stars observed at low or intermediate 
inclinations, the UV resonance ``wind" lines in the $\gamma$\,Cas spectrum 
indicate the presence of a variable high velocity wind. 
Henrichs, Hammerschlag-Hensberge, \& Howarth (1983) found that optical 
depths in the wind lines can vary by a factor of up to 20 over time.
The optical depths in the wind components increase by a mean factor of 2--3 
times from $V/R$ $<$1 to $V/R$ $>$1 phases.
Telting \& Kaper (1994) advanced a picture in which the wind ablates disk
matter, thereby focusing the wind and increasing column absorption at 
``blue" ($V/R$ $>$ 1) phases. Beyond this modulation, changes in wind line 
strength can vary by up to a factor of five in a few weeks. 

The cause of changes in the wind over these shorter timescales is a still 
unsolved mystery and is probably related to how Be stars change their local 
environment.  In a distinctly different scenario from the Telting-Kaper 
picture, Brown et al. (2008) have suggested that a global magnetic field 
from the star focuses the wind towards the disk, thus
producing a compressed and ``magnetically torqued disk." 
In such a picture one can imagine that time-dependent changes in the field
geometry would focus the wind to varying degrees across our line of sight at
different times, producing variations in the UV wind absorption components. 
This mechanism could not work for $\gamma$\,Cas globally
because a dipolar field would produce unobserved strong periodic emission and 
absorption variations over the rotation cycle, as observed only in Bp stars.
Still, it is conceivable that magnetic loops from chaotic surface fields
could divert wind flows on smaller scales.

H$\alpha$ emission and optical brightening as well as reddening episodes in
the optical records of $\gamma$\,Cas and other Be stars occur over timescales 
as short as a few weeks. 
In addition, Carciofi et al. (2012) have simulated a $\sim$100 day
dissipation of the disk of the Be star 28\,CMa with a viscous disk model
and concluded the high Shakura-Sunyaev viscosity implied, $\alpha$ 
= 1.0${\pm 0.2}$, requires 
that the mass injection rate into this disk exceeds the wind mass loss rate 
by about an order of magnitude. These considerations suggest that an 
unknown {\it deus ex machina} operates at irregular intervals that causes
discrete ejection events (``outbursts") in at least some Be stars.
In $\S$\ref{discss} we will return to this topic.  For now we will loosely 
use the term ``wind" to indicate any type of ejection mechanism, although they 
may well involve more than radiative and centrifugal forces.
\\

\noindent {\it Corotation and migrating subfeatures:~}  \\
 A number of observers have noted the existence of narrow absorptions called
``migrating subfeatures" ({\it msf}) that run blue to red across optical and 
UV lines of the $\gamma$\,Cas spectrum. 
(Yang et al. 1988, Horaguchi et al. 1994, Smith 1995, 
Smith \& Robinson 1999). In optical lines {\it msf} do not generally appear 
during every night of observation. When they do appear a new feature moves 
across a given point on the profile every 4-5 hours and at an acceleration 
rate of +95 km\,s$^{-1}$hr$^{-1}$. The {\it msf} have also been observed in 
spectral lines of the $\gamma$\,Cas analog HD\,110432 (SB06) and 
the magnetically active, rapidly rotating K0 star AB\,Dor (Collier
Cameron \& Robinson 1989) and are ascribed to forced-corotation of dense 
clouds over the star's surface by magnetic fields. Indeed, 
Donati et al. (1999) have confirmed the detection of a complex field
topology on the surface of AB\,Dor, reinforcing the corotation interpretation.

  The character of the spectroscopic {\it msf} makes it unlikely that they 
arise from nonradial pulsations (NRP). Recent photometric monitorings of 
Be stars have disclosed excited modes for which the periods tend to be 
clumped into noncoherent groups. For early-type Be stars they generally are 
excited with longer periods ($\sim$0.5 day$^{-1}$) than the times between 
passages of successive {\it msf} absorptions.\footnote{Exceptions to 
this statement are found in the spectral variations of
$\mu$\,Cen and $\delta$\,Sco, which can exhibit modes
with periods in this range in addition to others (Rivinius et al. 2003, 
Smith 1986). In these cases the appearance of broad sinuous NRP bumps on 
line profiles is very different than narrow {\it msf} depicted in the
references we have cited.} 
The observed NRP modes are
attributed mainly to rotational splittings of g-modes with $|m|$ $\approx$ 
2 (e.g., Walker et al. 2005, Dziembowski et al. 2007, Balona et al. 2011) 
or to inhomogeneities of orbiting disk matter (Balona 2009). Both their
short lifetimes and their appearance as high frequency ``wiggles" in the 
line profiles make the {\it msf} inconsistent with low-$|m|$ modes.

 The interpretation of forced corotation of clouds was also advanced to
explain dips in UV continuum light curves of $\gamma$\,Cas (Smith, 
Robinson, \& Hatzes 1998; ``SRH"). The span of these dips is too brief to 
be caused by advection of spots across the star's surface. 
They can be explained, however,
by the presence of translucent clouds corotating within a few tenths of 
the stellar radius of the surface. 
Also, the appearance of UV light curve dips correlates well with
the increases in X-ray flux over several hours, according to a
simultaneous X-ray/UV campaign described by SRC.

\section{Origin of the X-ray emissions}
 
The ultimate goal of our ongoing studies of this topic is to define the 
mechanism responsible for the unique emission of the hard and secondary 
soft X-ray emissions that set the $\gamma$\,Cas variables apart from the 
emissions of other high-mass stars. Most recent suggestions for this
origin have centered around two competing ideas.

According to the first idea 
the hard X-rays result from the conversion of gravitational
energy from accretion of the Be star's disk onto the surface of a
degenerate companion (white dwarf or neutron star). The primary rationale
for this idea is the resounding success in explaining X-ray active Be
stars in the Galaxy and Magellanic Clouds. Indeed, most of these X-ray
sources are known to be binaries because of the detection of X-ray pulses
due to a neutron star. In the case of the $\gamma$\,Cas stars no such
pulses are known, and the direct detection of binarity through radial
velocity variations has so far been confined to $\gamma$\,Cas itself.
As already indicated, the near-circular orbit of this system hints that
the secondary is not a neutron star. 
In this connection it may be significant that two members of the $\gamma$\,Cas 
group, ``Star 9" of NGC\,6649 and HD\,119682 seem to be blue stragglers 
of clusters having ages of about 50\,Myr and 40\,Myr, respectively
(Marco, Negueruela, \& Motch 2007, Marco et al. 2009). Moreover, 
HD\,110432 may be a member of NGC\,4609 (age of 60\,Myr; Feinstein
\& Marraco 1979), and if so would be a straggler.
Thus, while unproven, it is conceivable 
that all $\gamma$\,Cas stars are blue stragglers. This circumstance could 
mean either that the current degenerate secondary star is the site of the
X-ray emission or that it simply provides a mechanism to spin up the outer 
layers of the Be star to near criticality, thereby somehow triggering a 
mechanism for the production of the X-ray emissions.

  Another argument for tying the X-ray emission to white dwarfs is
the multi-thermal nature of the X-ray plasma observed in some cataclysmic
variables, particularly the magnetic ``polars" (e.g., Mukai et al. 
2003). However, upon close inspection there are also significant differences
between the X-ray characteristics of these objects and $\gamma$\,Cas. These 
include in the former case a general lack of ubiquitous flaring confined
to the hot X-ray regions and their exhibiting no differences in the 
Fe abundance determined from L and K shell ions (e.g., Itoh et al. 2006).  
In addition, even in their high states the hard components of polars have
lower fluxes, L$_x$ $\sim$10$^{30}$-10$^{32}$ erg\,s$^{-1}$ (e.g., Watson 
\& King 1987) than $\gamma$\,Cas . If accreting white 
dwarfs are the source of the X-rays, it is likely 
that they would represent a new subclass of such objects.

 A very different concept to explain the X-ray flux of $\gamma$\,Cas is
called the ``magnetic star-disk interaction" model and was 
proposed by RSH. These authors pointed to the X-ray flares noted above,
the correlation of rapid
variations of X-ray flux and UV and optical diagnostics associated with
matter in the proximity of a Be star, some of which were noted above, 
as well as the correlation of the long X-ray and optical cycles. 
In this picture magnetic lines of force
from active regions on the rapidly rotating Be star and its slower rotating 
disk occasionally get entangled. The result is that they stretch, 
sever, and upon reconnection fling high energy particles in various 
directions including toward the Be star. The impacts of these particles
on random points on its surface heat photospheric plasma, and they 
become visible as ``flares" - though not flares in the solar sense of
releasing local magnetic energy.
This matter expands to lower density regions
above the flare site and continues to glow as ``basal" emission. The
observation of X-ray flux ``lulls" over several epochs suggests that
this does not happen perfectly continuously but instead can be subject
to relaxation cycles. 

  The heart of the evidence for the Be star-disk magnetic interaction 
picture lies in observations of rapid correlations between X-ray flux 
and fluxes in other wavelength regimes and, secondarily, on inferences of 
magnetic activity on the star's surface. The first well documented
example of such a correlation was the report by Slettebak \& Snow
(1978) of an H$\alpha$ emission
``flare" that occurred at the same time as short-lived occurrences
of emissions in scanned components of the Mg\,II and
Si\,IV lines by the {\it Copernicus} UV-scanning spectroscopic satellite.
This could not be dismissed as a (double) instrumental ``fluke" because {\it
Copernicus,} using its coaligned X-ray telescope,  also observed an X-ray
during this time (Peters 1982). On another occasion, Smith (1995) observed 
a brief ``emission flare" in the He\,I 6678\,\AA\ line,
although no other wavelength regions were monitored.  

 To pursue this evidence SRC mounted a 21 hour simultaneous
campaign using  the {\it Rossi X-ray Timing Explorer (RXTE)}
and the Hubble {\it Goddard High Resolution Spectrograph} in 1996 March. 
They identified several correlations between X-ray and UV continuum flux
as well as various spectral lines expected to be formed in the neighborhood 
of a B0.5 star (e.g., Smith \& Robinson 2003). These variations are the 
strongest single set of arguments that
the X-rays are excited near the surface of the Be star.

The correct explanation for the production of the hard X-rays will 
have interesting ramifications for our understanding evolutionary and
possibly the angular momentum history of the X-ray emitting star. 
The elucidation of this true mechanism will require the pulling together 
of diverse properties of the Be star, its binary orbit, and circumstellar 
matter from observations at several wavelength regimes over as 
contemporaneous a time interval as possible.
 The following describes a comprehensive effort to this end.

\section{Observations}

We first outline the observational techniques used in this paper.
We begin by summarizing the
LBOI results and discussing results from automated photometry 
and H$\alpha$ line spectroscopy.

\subsection{The CHARA Long Baseline Optical Interferometer}

  Our LBOI observations were conducted in the infrared using 
Georgia State University's Center for High Angular Resolution 
Astronomy (CHARA) array facility on Mt. Wilson. 
CHARA is an interferometric array of six 1-meter telescopes 
configured in a Y-shaped pattern with baselines of up to 330 m 
for maximum spatial resolution. Our observations utilized 
two new-generation instruments for imaging of $\gamma$\,Cas's 
decretion disk: the Michigan Infrared Combiner {\it (MIRC)} 
in the H-band (1.64 $\mu$m) and the Visible spEctroGraph and
polArimeter {\it (VEGA),} which became operational in 2009 
(Mourard et al. 2009). {\it VEGA} observations were conducted 
in H$\alpha$ and also near this line in $\sim$200\,\AA\ 
regions centered at 6460\,\AA, 6690\,\AA, and 6728\,\AA.~ 
In the first case the results from the three wavelength settings
were averaged together and we call the result the ``R"-band measures.
The H$\alpha$ line fluxes were determined
by stepping in 3\,\AA\ steps across the 6550--6575\,\AA\ region.
The observations were executed in order to describe the geometry of
the Be disk. The {\it VEGA} kinematical program was conducted
to test the disk rotation-radius relation. These observations 
are summarized in the following and are detailed in Stee et al.
(2012; ``Paper\,2").

The {\it MIRC} instrument was used in a 4-telescope arrangement, resulting
in the simultaneous measurement of 6 visibilities, 4 closure phases and
4 triple amplitudes.  These observations used a broadband H filter along
with a prism to disperse the light into 8 spectral channels (R$\sim$45). 
This light
is refocused to produce interference fringes received by a ``PICNIC" camera
designed to maximize rapid data readout and minimize fringe
decoherence due to atmospheric seeing.  Further details on this
electronic and readout system may be found in Monnier et al.
(2004, 2008), Pedretti et al. (2009), and Che et al. (2010).
The maximum angular resolution $\frac{\lambda}{2B}$ in
the H band using the {\it MIRC} at CHARA is 0.4 milliarcsec, or
coincidentally about the expected angular size of $\gamma$\,Cas, the Be star.
The observations for this study were carried out on 
2010 Aug. 5, Aug. 8, Sept. 24, and Nov. 4-5. This range of dates 
includes the range of the {\it XMM-Newton} OBS 3 and 4 referred
to below and the next two months.
These observations were followed almost immediately by {\it VEGA} 
observations. The MIRC baselines ranged 
from 34\,m (S1S2) to the maximum 330\,m (S1E1).

   The {\it VEGA} instrument is different from {\it MIRC} 
in working with visible wavelength (4500-8500\,\AA) light at 
``low, medium, or high" dispersions (1,700, 5,000, 30,000) but 
enables the combining of light from up to four CHARA telescopes 
and measurements of visibility modulus, phases, and phase closures
as a function of baseline and wavelength. Its angular resolution 
is 0.3 milliarcsecond in the visible. 
The addition of a Wollaston prism allows the study of polarization 
across velocity dispersed systems, such as the circumstellar disk of 
$\gamma$\,Cas. The performance of the basic instrument is
summarized by Mourard et al. (2011).

 Begun in 2008, the {\it VEGA} observations of 
$\gamma$\,Cas summarized in this paper and as noted in Paper\,2.
As noted in the latter, half of them were contemporaneous with 
the {\it XMM-Newton} observations, and nost of these
measurements were made over a variety of configurations.
A limitation of our {\it VEGA} dataset is that the observations
were clustered near the same minimum baseline ($\sim$30\,m) in 
the South (S1S2) alignment. The W1W2 baseline was more than
adequate to resolve the disk in the East-West direction on the sky. 
However, this mismatch in configurations limited our ability to verify 
the presence of features in the South-North direction on the disk, 
even though departures from disk axisymmetry have been observed with 
earlier generations of equipment (e.g., Berio et al. 1999).
Also, we were unable to get a complete set of phase closures at multiple
times.  This means that we could not
determine time-dependent differences during the 2008-2010 interval.

\subsection{The Automated Photometric Telescope Observations}

  Our optical continuum observations of $\gamma$\,Cas were made in the 
standard Johnson $V$ and $B$ band filters on every available photometric 
night (sometimes several times per night) using the 0.4 meter T3
Automated Photometric Telescope (APT) located at Fairborn Observatory
in the Patagonia mountains in southern Arizona. Our monitoring of this 
star began in 1997 September and has continued every year since, with
an observing season lasting from September into February and restarting 
again briefly in mid-June, weather conditions permitting.  We lose about
two months of the  $\gamma$\,Cas observing season due to monsoon rains.

Full details concerning our APT observational program on $\gamma$\,Cas
can be found in RSH and SHV. 
Briefly, the observations are made by cycling the automated telescope
between the program star and the nearby stars HR\,6210 (F6\,V) and HD\,5395 
(G8\,IIIb), which are used as comparison and check stars, respectively.
The APT executes each observing sequence as follows: K,S,C,V,C,V,C,V,C,S,K, 
where K, C, and V, are the check, comparison, and program stars, respectively,
and S is a sky observation.  Each complete sequence is termed a ``group
observation" or simply a group. All integration times were 10\,s, and all
measurements were made through a 3.8~mag neutral density filter to attenuate 
the counts and avoid large deadtime corrections.  The program star is thus 
bracketed by the comparison star on both sides three times during each 
group measurement at a cadence of 8 minutes.  
Group mean $V-C$ and $K-C$
differential magnitudes are computed and standardized with nightly 
extinction and yearly mean transformation coefficients determined from 
nightly observations of a network of standard stars.  
Typical {\it rms} errors 
for a single observation, as measured from pairs of constant stars, are 
$\pm{0.0030-0.0035}$~mag.  The {\it rms} scatter of the seasonal mean 
$K-C$ differential magnitudes, inferred from differences of the
40th and 60th percentile of fluxes during a season (a reliable internal
metric) is $\sim{\pm 0.003}$ mag., indicating good long-term stability. 
By comparison, the 
corresponding figure during the 2010 season was twice as large, 
${\pm 0.007}$ mag. This was also near the value of the formal {\it rms}.
These larger values are largely due to the quasi-secular brightening during
this time.

%The $V-C$ differential magnitudes were converted to apparent magnitudes of
%$\gamma$\,Cas by assuming apparent magnitudes m$_V$ = 5.84 and m$_B$ = 6.40 
%for the comparison star. 

\subsection{Optical wavelength spectra of $\gamma$\,Cas}

  To support this program we made use of 306 echelle spectra of 
$\gamma$\,Cas taken in a long-term monitoring program with the 
spectrograph fiber-fed from the Cassegrain focus of the 
1.06-meter telescope at Ritter Observatory (administered by the 
University of Toledo) in 2009 December and the second half of 2010.  
The spectra were obtained using a new Spectral Instruments, Series 
\#600 camera configured in an echelle format. This 
%pixels in an echelle spectral format on a 4K$\times$4K pixel CCD,
resulted in a resolution of 27,000.  We used three echelle orders
to extract the H$\alpha$ spectrum and background, extending over
the range 6285--6834\,\AA.~ Exposures lasting a
few minutes gave spectra with a signal to noise ratio of $\approx$100. 

 Our dataset includes H$\alpha$ spectra by E. Pollmann
who used a cooled CCD camera with 9$\times$9 $\mu$m pixels attached to  
the Cassegrain focus of his 35\,cm Schmidt telescope in Leverkusen.
A dispersion of 22.2 \AA\,mm$^{-1}$ made for a spectral resolution of
 16,400  over the range 6520-6700\,\AA.  In Hamburg R. B\"ucke used a 
20\,cm Dobson Telescope fiber coupled to a spectrograph contain a cooled 
CCD camera with 18$\times$200 $\mu$m pixels.  This system provides a 
dispersion of 0.56\,\AA\,pixel$^{-1}$ and a spectral resolution of 3,400 
extending over a range 5900--6700\,\AA.
These instruments have been used in stable configurations for several years.

\subsection{XMM-Newton Observations}

   We conducted four {\it XMM-Newton} observations under the auspices of 
Guest Investigator cycle A09.
These were carried out on 2010 July 7 (ObsID 0651670201; referred to here as 
OBS1), July 24 (0651670301; OBS2), August 2 (0651670401; OBS3), and August
24 (0651670501; OBS4), beginning on MJD dates 55384.30, 55401.15, 55410.77, 
and 55428.09, respectively; {\it XMM-Newton} prefers the MJD time system. 
% MIDEXP TIMES: 55384.503, 55401.332, 55410.974, 55428.349
%All observations were made during intervals free of
The observations OBS1 to OBS4 had total effective exposure times 
of 17.5, 15.7, 17.5 and 22.4 ks for the $pn$ detector, 
respectively, 0.2\,ks longer times for the {\it MOS1} and {\it MOS2} cameras, 
and 0.4\,ks for the {\it RGS} cameras. High energy-sensitive
data were taken through the thick filter of the {\it European Photon Imaging
Camera (EPIC)} camera operating in small window mode. 
To avoid high noise contributions from soft energies, our timing 
analyses were restricted to {\it EPIC} data in the 0.8--10\,keV energy band. 
Our spectral analysis made use of the complementary 
information of the medium resolution {\it EPIC} spectra between 0.2 
and 12 keV and high resolution {\it RGS} spectra
from 0.3 and 2 keV. 
Data reduction, spectral analysis and part of the timing analysis were 
carried out using SAS v10.0.0, XSPEC, and Xronos package, respectively.
See http://heasarc.nasa.gov/xanadu/spec/ for a description of these packages.

 Altogether we used five {\it XMM-Newton} cameras in our observations.
Of these the {\it EPIC} detectors ({\it pn, MOS1,} and {\it MOS2}) were 
especially important for the evaluation of the hot plasma component.
The {\it RGS1} and {\it RGS2} cameras were more important for assessing
the warm and cool plasma components defined below, and exclusively in
determining spectral line attributes for lines above 10\,\AA.
The optical monitor was turned off during our observation.

\section{Results of optical investigations}

\subsection{Fitting of CHARA visibilities of the $\gamma$\,Cas system}
\label{lboi}

   In this section we summarize the interferometric results obtained in
our collaborative program.  

We were able to resolve the disk diameter in both H-band and R-band  using 
LITpro model programs (for details see http://www.jmmc.fr/litpro\_page.htm).
We fit the MIRC data with two three-component models, that is with 
a uniform brightness star disk, a Be disk and a uniform ``background" 
comprising 12--14 \% of the total flux.  Such a component is not 
the norm for Be-disk descriptions.  Physically it could be due to the 
presence of previously ejected matter from the star that is now extended 
beyond the disk. In the first model the Be disk
is assumed to have a Gaussian profile with distance from the center of the 
star. In the second model the disk model is an elliptical ring of equal 
surface brightness, with a brightness cutoff at the inner and outer edges. 
The motivation for the ring model was to determine whether there is a 
star-disk gap that might arise from solid body rotation of clouds within 
the radius.  However, we were unable to get a satisfactory fit to the 
VEGA data using the ring model.  Table\,1 shows the results.
The angular values in milliarcsec here are converted
to linear dimensions and represented in italics, 
assuming the distance and value of R$_*$ given in $\S$\ref{intro}.

Using these models with the MIRC data, we resolved a
flattened circumstellar disk with elongation ratios of 0.82/0.62 =
1.33 (Gaussian) and 0.85/0.61 = 1.39 (ring).
The VEGA R-band data led to an elongation of 0.76/0.56 = 1.36 % \,${\pm .14}$
The MIRC values give an inclination $i$$\approx$42$^o$ and 44$^\circ$, 
respectively.  
The VEGA R-band results are essentially the same, 
i$\approx$42$^{\circ}$${\pm 3}$$^{\circ}$. 
Likewise, the position angle derived from the VEGA is 
19$^{\circ}{\pm 5}^{\circ}$ and thus consistent with the MIRC value.
For the MIRC solution the ring radius extends from 1.4\,R$_*$ to 2.0\,R$_*$,
but the error bars are large. 

As Table\,1 shows, we were unable to get a satisfactory solution for the 
{\it VEGA} R-band data with a ring model. 

%%%%

\begin{table*} \caption{Relevant star and Be disk fitting parameters from LBOI}        
\label{tbl:lboi}        \centering           \begin{tabular}{ccc}

\hline
\hline\\[-2.2ex]

MIRC DATA: IR H-band  &         &     \\
%\tableline

\hline
PARAMETER             &  GAUSSIAN DISK MODEL  &     RING DISK MODEL \\
\hline
Star Diameter (mas)   &  0.44${\pm 0.03}$ &   0.44${\pm 0.03}$ \\
     &         &      \\
Inclination angle (i) & 42$\pm 1\degr$ &44 $\pm 1\degr$\\
(thin disk or ring) &  & \\
Minor disk axis (mas)           &   0.62${\pm 0.08}$  &       ...   \\
{\it (Linear, in 2R$_{\odot}$)}  &   {\it 1.4}   &       ...       \\
Major disk axis (mas)           &   0.82${\pm 0.08}$    &        ...     \\
{\it(Linear, in 2R$_{\odot}$)}  &  {\it 1.86}                & ...      \\
Minor ring inner diam. (mas)  &   n/a                &    0.60${\pm 0.33}$  \\
(major axis)            &       &\\
Outer ring diameter (mas)        &   n/a                &    0.85${\pm 0.45}$  \\
along the major axis            &       &\\
Flattening ratio        & 1.33${\pm 0.08}$ & 1.39${\pm 0.08}$\\
Position Angle              &   12${\pm 9}$  &    12${\pm 9}$ \\
(E from N in degrees)       &                      &                      \\
                            &                      &                      \\
Star flux (\%)         &          35${\pm 1}$      &      45${\pm 6}$      \\
Disk/ring flux (\%)    &         53${\pm 2}$      &      41${\pm 6}$      \\
Background flux (\%)   &        12${\pm 1}$      &      14${\pm 1}$      \\
$\chi^{2}$         &              3.96        &            3.95            \\
                            &                      &                      \\
\hline
VEGA: near-H$\alpha$ ``R"-band  &                 &                      \\
\hline
PARAMETER             &  GAUSSIAN DISK MODEL  &     RING DISK MODEL \\
\hline
Star diameter (mas)      &     0.44${\pm 0.03}$   &     (no fit)         \\
                            &                      &                      \\
Inclination angle (i) & 42$\pm 3\degr$ &n/a\\
(thin disk) &  & \\
Minor disk axis (mas)             &  0.56${\pm 0.05}$   &       n/a           \\
Major disk axis (mas)             &  0.76${\pm 0.05}$   &       n/a          \\
{\it Major axis (in 2R$_{\odot}$)} &  {\it 1.72}   &                     \\
Flattening ratio        & 1.36${\pm 0.08}$ &  n/a\\
Position Angle              &   19${\pm 5}$        &        n/a          \\
(E from N in degrees)       &                      &                      \\
                          &                        &                      \\
Star flux  (\%)          &      55${\pm 2}$        &         n/a         \\
Disk flux  (\&)          &      45${\pm 2}$        &        n/a          \\
$\chi^2$                 &      1.95               &        large        \\
\hline

\hline                                  \end{tabular}
%\begin{list}{}{}
%\end{list}
\end{table*}

  We also used the VEGA system to measure fluxes in the H$\alpha$ line
both spatially and kinematically.  In this analysis we adopted
the major axis and inclination angle from the R band analysis
outlined above. Our best fitting major axis parameter in H$\alpha$ flux 
is 4.4${\pm 0.4}$ mas.  The position angle was measured as 
20${\pm 15}$$^{\circ}$.
Assuming the previous parameters, we find a Be disk
radius of 10${\pm 1}$\,R$_{*}$. We point out that at 10\,R$_*$ the
VEGA disk radius is nearly half the Be star's Roche lobe radius (of
some 21.4 R$_*$; Gies et al. 2007). 
% Gies et al. 2007 find a orbital separation of 37R*, given R* = 10R_s
We find from Okazaki \& Negueruela (2001)
that the most likely 3:1 resonance truncation radius of the disk due to the 
companion in a circular orbit is about 0.4 of the orbital separation, or 
$\approx{\frac 34}$ of the Roche lobe radius. Our VEGA radius does not yet 
reach this extent, but one may hope more sensitive future observations,
particularly in the far-IR, that could test for the presence of a truncation 
edge.

 Our kinematic observations for H$\alpha$ were designed
to determine the exponent value $\beta$ in the velocity
distance relation in the disk, V$_{\phi}$ = (r/R$_{*})^\beta$. 
We obtained a value of $\beta$ = 0.5${\pm 0.1}$, or exactly
the Keplerian rotation result. Solid body rotation,
for which $\beta$ = -1, is therefore (unsurprisingly) ruled out for the
whole disk.  The confluence of Keplerian rotation with corotating 
 clouds over the star's surface suggests a complicated geometry
for the CS matter.

\subsection{Optical photometry and spectroscopy}

\subsubsection{Combining the APT light curve and H$\alpha$ measures }

\begin{figure}
   \centering
   \includegraphics[width=6cm,angle=90]{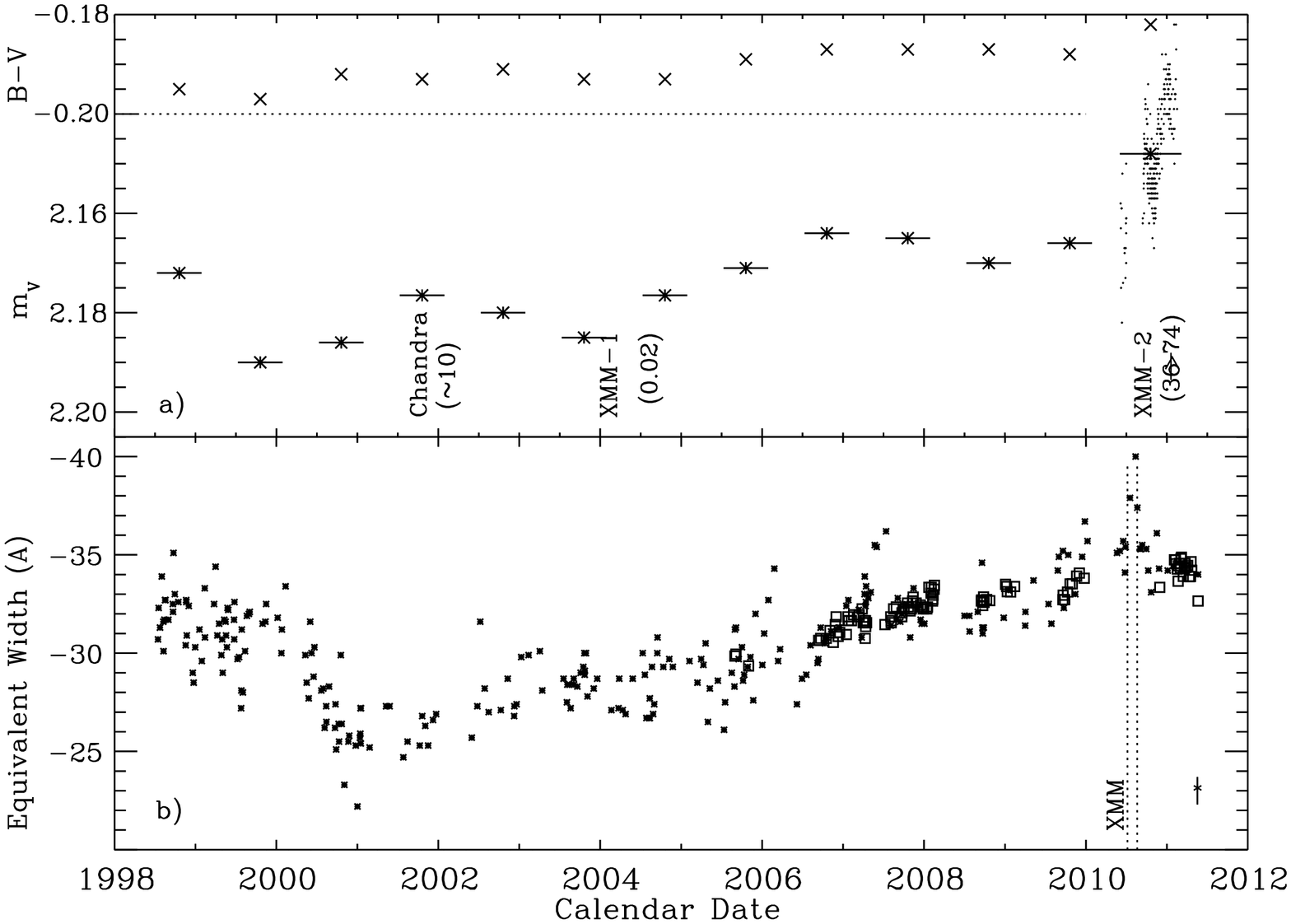}
      \caption{
(a): APT $V$ (m$_V$) magnitudes and $(B-V)$ color of $\gamma$\,Cas 
over the 1998--2010 seasons of coverage; annual averages are shown.
Error bars 
on these magnitudes (marked as * and X, respectively)  are much smaller than 
the symbols, as each point represents a few hundred individual observations. 
Horizontal bars denote the coverage in time for each season's average. 
Small points indicate each m$_V$ value measured in the 2010-2011 season.
Annotations at the bottom give the column densities n$_{H_b}$ in units of
10$^{22}$ cm$^{-2}$ responsible for the attenuations of soft X-rays in
high dispersion X-ray observations of 2001, 2004, and 2010. 
(b): The H$\alpha$ emission equivalent widths (EW) in \AA ngstroms obtained by 
Pollmann and B\"ucke (star and square symbols, respectively over the same time 
as panel (a)). A displacement of -1.0 \AA\ was required to bring the 
the mean B\"ucke EW to the Pollmann mean. 
The vertical dotted lines correspond to times of the first and last
{\it XMM-Newton} observations in 2010.  
Note the rapid H$\alpha$ emission increase from 2002 on,
and also a similar spike in 2007 as observed in 2010. Both events coincide 
% with m$_V$ brightenings in blow-ups of panel (a).
with m$_V$ brightenings when panel a is plotted in high time resolution.
              }
         \label{ltcrv}
   \end{figure}

Figure\,\ref{ltcrv}a plots the seasonal mean Johnson $V$ magnitudes and
% Fig 1 introduced here
mean $(B-V)$ color indices from the T3 APT for the 13 observing seasons 
from 1998--2010.\footnote{We define the ``1998" season as 
the interval from 1998 June--1999 February. Note that we omit the 1997 
observing season because a neutral-density filter change in 1998 resulted 
in new magnitude zeropoints from 1998 onwards.  Analysis of the 
variability in 1997 was published by SHV.  Errors in the seasonal means 
are very small and are due mainly to incomplete averaging of the long 
cycles and/or slopes in the magnitude. } 

Notice a series of long-term erratic undulations over timescales of 
2--3 years. This variability pattern probably results from 
structural changes in the disk, but the details of this are unclear.
We can shed some light on this by coplotting the H$\alpha$ measures
in Figure\,\ref{ltcrv}b. 
Although the light curve shows highly statistical fluctuations from year 
to year, over the entire timespan the slopes of both H$\alpha$ 
equivalent width and $V$-band light curves are positive - that is, 
the activity as defined  by H$\alpha$ emission, magnitude brightening, 
and $B-V$ color have increased more or less together. 
Thus the H$\alpha$ emission from the CS disk correlates with optical
continuum brightening.

Inspection of the two curves in Fig.\,\ref{ltcrv} shows that second order 
deviations in their behaviors exist. These include a delay in the decrease 
in the H$\alpha$ emission index over the Reduced Heliocentric Julian 
Date (RHJD) range 51040-51080, absences of rises 
in the index at RHJD 52000--52400 and at 54000--54200, and absence of a 
rapid upturn at the end of the data sequence.
Such deviations can arise because of differences in the optical thickness
in the $V$-band continuum and the H$\alpha$ line core. The optical depth in
the H$\alpha$ line is many times more optically thick than the $V$-band 
continuum (see e.g., Carciofi 2011) and therefore the point at which 
$\tau_{disk}$ = 1 lies further from the star. Consequently, the H$\alpha$ 
emphasizes disk changes at greater radius than the $V$ continuum does. 
%Also,inemail: (Stee 2001) (e.g., Hummel \& Vrancken 1995, Stee \& Bittar 2001)

\subsubsection{A revised radial velocity solution for the binary orbit}

We measured the radial velocities for the  H\,$\alpha$ spectra from Ritter 
Observatory by computing bisectors of the 
emission line components.  For this purpose we used the method of 
Shafter, Szkody, \& Thorstenson (1986), which has special 
application to measurements of shifts of asymmetric line profiles.
The bisector position is determined by creating a function composed
of a negative-valued Gaussian (offset by $\triangle V
=-300$ km~s$^{-1}$) and a positive-valued Gaussian
(offset by $\triangle V=+300$ km~s$^{-1}$); both
Gaussians have a width FHWM = 85 km~s$^{-1}$.
This function is then cross-correlated with the
observed emission lines (whose baseline is reset to zero),
and the zero-crossing of the resulting cross-correlation
function yields the bisector position.  We measured 198
velocities from Ritter Observatory that span from
2000 January 28 to 2010 November 11. 
Also, one of us (RB)  obtained 103 new measurements over the interval
2006 September 20 through 2011 April 22 from the Hamburg Dobson Telescope 
system.  These were measured by fitting a Gaussian function to the profile. 
Finally, we combined the Ritter and RB velocities. 
The weights of the Ritter data were determined from the mean 
continuum signal-to-noise ratio in the spectra.  Because the RB spectra 
were measured from lower resolution spectra, we assigned them a higher 
velocity error of ${\pm 3.16}$ km\,s$^{-1}$. The RB origin can be
found from this assignment in Table\,2.
To bring the Ritter and DB velocities to a common mean of zero, we added
5.73 km\,s$^{-1}$ and 9.95 km\,s$^{-1}$ to the Ritter and RB velocities,
respectively.  The results are shown in Table\,2 as a stub (dates in 
Reduced Heliocentric Julian Dates). The full data list of 301 radial 
velocities is given in the 
electronic version of this paper.  Column 3 is the measurement error 
in km\,s$^{-1}$; column 4 is the
``O - C" (Observed minus Computed) differences; column 5 the orbital
binary phase, and column 6 is the RV source (Ritter or RB).

\begin{table}   % add * so \begin{table*} for 2 col table; \end{table*} too
\caption{Ritter \& Dobson/RB radial velocities (stub)}              
\label{tbl:rvel}        \centering           
\begin{tabular}{c|c|c|c|c|c}             

\hline                    
\hline\\[-2.2ex]
RHJD & $\triangle RV$ & Error & O-C & Phase & Source \\ 
\hline                    

51571.510 &  1.63 & 0.92 & -2.16 & 0.00 & Ritter \\
51586.504 &  3.04 & 1.32 & -0.30 & 0.08 & Ritter \\
51608.527 &  0.38 & 1.31 & -1.09 & 0.19 & Ritter \\
51609.516 & -0.22 & 1.69 & -1.59 & 0.19 & Ritter \\
51777.805 &  3.80 & 5.81 &  0.04 & 0.02 & Ritter \\
51794.820 &  2.87 & 1.64 & -0.17 & 0.10 & Ritter \\
51800.800 &  0.31 & 1.19 & -2.26 & 0.13 & Ritter \\

\hline                                  
\end{tabular} 
\end{table}

%%%%%
\begin{table*} \caption{Orbital Solution for $\gamma$ Cas.}              
\label{tbl:orb}        \centering           \begin{tabular}{l c c c c}            

\hline                    
\hline\\[-2.2ex]

Parameter &   Units        &    1    &       2 (adopted) \\ 
\hline
$K$       &  (km s$^{-1}$)  &  $3.80 \pm 0.08$  & $3.79 \pm 0.08$   \\
$e$       &                 &  $0.03 \pm 0.02$  &  ... \\
$\omega$  & (degrees)       &  $198 \pm 29$  &       ...  \\
$T$    & (RHJD) & $55549.48  \pm 16.32$ & ... \\
$T_{\rm RV max}$ & (RHJD) & $55436.47 \pm 0.64$ &  $55438.1045 \pm 1.35$  \\
$P$       &       (days)     & $203.590 \pm 0.21$  & $203.555 \pm 0.21$ \\
$f(M)$    &  ($M_{\odot}$)  & $(0.116 \pm 0.008)\times 10^{-2}$ & $(0.115 \pm 0.007)\times 10^{-2}$ \\
$a_1 \sin i$  & $(R_{\odot})$  & $15.28 \pm 0.33$     & $15.23 \pm 0.33$    \\ 
$rms $      & (km s$^{-1}$)   &  1.03                &  1.03     \\
\hline                                  
\end{tabular} 
\end{table*}
%%%%%

\begin{figure}
   \centering
   \includegraphics[width=6cm,angle=90]{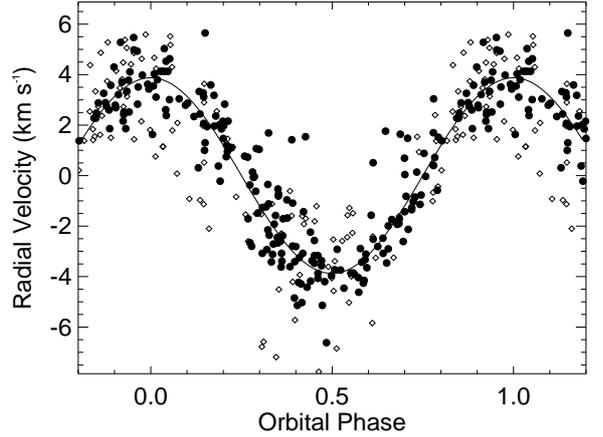}
      \caption{
      The adopted radial velocity solution from our Orbit 2 model (see Table 3).
Solid dots and open triangles are measures obtained from Ritter Observatory and
RB's Hamburg system, respectively. The weights to the sinusoidal solution are 
lower for the open symbols. Table 2 gives errors for all observations.
              }
         \label{rvfig}
   \end{figure}

 In order to construct a new orbital solution, we first examined these new 
data and found evidence for long-term trends introduced by including pre-2000 
radial velocities. 
%In their study of $\gamma$\,Cas's radial velocities
%Miroshnichenko et al (2002) also used pre-2000 data. 
We did not include these early data because the methods for pre-whitening 
could introduce new artifacts in the analysis of the orbital elements. 
%In order to fit the
%data better, we derived an average radial velocity for both data sets, 
%and shifted both data sets to a systematic zero velocity.
We made an error-weighted fit to the velocities to solve for orbital 
elements using the program of Morbey \& Brosterhus (1974).
We solved for the orbit by allowing the eccentricity $e$ as well 
as other orbital parameters to float freely. 
This solution is represented as Orbit\,1 in Table\,3. 
Discovering that this eccentricity is at most no more than marginally 
significant (e.g., from testing the probability index $p_1$ suggested
by Lucy 2005), we ultimately adopted the circular solution with the same
period.  This is given Orbit\,2 in our table. We depict the folded data
to the Orbit\,2 solution as a 203.555 day sinusoid in Figure\,\ref{rvfig}.
Phase zero for Orbit 2 is referred to the maximum velocity.
For Orbit 1 we adopt the time of periastron passage.

Our derived orbital parameters for $\gamma$\,Cas differ slightly from the 
discovery analyses of Harmanec et al. (2001) and Miroshnichenko et al. 
(2002).  The first study found $e$ = 0.26 and P$_{orb}$ = 203.59 days, 
and the second $e$ = 0.0 and P$_{orb}$ = 205.50 days.  Essentially, 
our new period solution agrees with Harmanec et al. but our
eccentricity value agrees with Miroshnichenko et al.
Shortly before submission of our results for publication, we received
a preprint from J. Nemravov\'a and collaborators
describing a radial velocity study covering
nearly 17 years of the star's velocity history (Nemravov\'a et al. 2012).
Their results and ephemeris are substantially the same as our own. Their
estimate of the period, 203.52${\pm 0.08}$ days, is well within the
formal errors of our two analyses.
The same comment applies to the errors in T$_0$ and phases. Therefore,
we believe a synthesis of our works leads to a definitive orbital parameters.

 From the parameters in Table\,3, if we assume a mass of 15M$_{\odot}$ 
for the primary and an inclination of 45$^{\circ}$ and assume that the 
binary orbit and Be disks are coplanar, we may adopt the mass function
of 1.15$\times$10$^{-3}$M$_{\odot}$ to find a most probable 
secondary mass of 0.8\,M$_{\odot}$. Assuming reasonable errors 
for the primary mass (${\pm 1}$\,M$_{\odot}$), an inclination of 
45$^{\circ}$ (${\pm 6}$$^{\circ}$), and K$_1$ (${\pm 0.1}$\,km\,s$^{-1}$), we 
estimate a generous error of ${\pm 0.4}$ M$_{\odot}$ for the secondary. 
This estimate supports the possibility that the secondary could be a late
type dwarf, white dwarf, or a sdO star, but probably not a neutron star.
This justifies the omission of this star from adding an additional
optical stellar component in our LBOI models.

  If we adopt the Orbit\,2 determination in Table\,3, our orbital 
phases corresponding to the {\it XMM-Newton} observations discussed 
below are $\phi_{orb}$ = 0.74, 0.82, 0.87, and 0.96 (where again 
$\phi$$_{orb}$ = 0.0 is defined relative to the epoch of maximum RV). 
The Nemravov\'a et al. RV orbital ephemeris gives phases for these 
observations of 0.73, 0.81, 0.86, and 0.95, i.e., within 0.01 of our 
values. As for our solution, Nemravov\'a preferred a circular solution 
to even small values of the eccentricity. In considering the relative 
weighting of these two RV studies, we can take the errors at face 
value and double weight the Nemravov\'a et al.  solution relative to 
our own. This gives a period of 203.53${\pm 0.10}$ days.

To summarize, the first of the 2010 {\it XMM-Newton} observations was 
conducted just as the Be star was in inferior conjunction with respect to 
its secondary, similar to the phase of the 2001 {\it Chandra} observation. 
The last of the four observations coincided with the Be star 
essentially at quadrature phase. With this emphemeris one finds that
the phase of the 2004 {\it XMM-Newton} observation was 0.27.

\section{Analysis of XMM-Newton Observations}

  Herein we discuss four high-quality {\it XMM-Newton} exposures of 
$\gamma$\,Cas in 2010, dubbed OBS1-4 above. 
The star was previously observed in high resolution
on two occasions: in August, 2001, by the High Transmission Energy
Grating of Chandra (HEG/MEG) and in February, 2004,
by {\it XMM} (RGS) (Smith et al. 2004, Lopes de Oliveira et al. 2010). We
can combine the photometric and spectroscopic results to get a sense 
of epochal differences of the conditions responsible for X-ray emission.

\subsection{XMM-Newton light curve variations}
\label{xrlite}

 We began the analysis by extracting light curves binned to 20 seconds 
for hard- (4.5-10\,keV), soft- (0.3-1\,keV), and the total flux over 
0.3-10\,keV for each of the four observations.
The median fluxes of three of the light curves extracted from the $pn$ 
detector (0.3-10 keV) are about the same, about 25.0 counts\,s$^{-1}$. 
The mean flux of the third observation at MJD\,55410 is 47\% higher than the 
mean of the other three, among which the difference is only ${\pm 7}$\%. 
{\bf Given the assumed distance of 168\,pc, this means that the L$_x$'s for 
the observations lie in the range 8--13$\times$10$^{32}$ erg\,s$^{-1}$.} 
This range is somewhat higher than the range 
{\bf 3--9$\times$10$^{32}$ erg\,s$^{-1}$} 
from RSH using the {\it RXTE} several years earlier. 
% NEW: #s in previous two lines have to be scaled by *0.80X 

\begin{figure}
   \centering
   \includegraphics[width=6cm,angle=90]{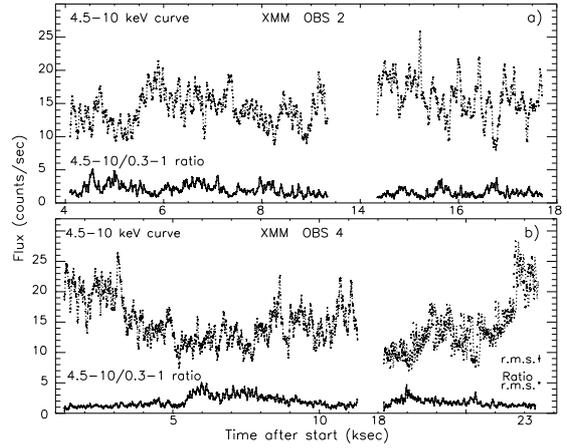}
      \caption{
      Two stretches of {\it XMM-Newton} light curves and color ratios for OBS2 
and OBS4. The light curves are binned to 20 seconds  and 
smoothed over two points. The {\it rms} errors are indicated taken from 
point to point fluctuations, with the same smoothing. The color ratios 
have been scaled by 5$\times$ for visual clarity. Occasional flare aggregates
are obvious in the light curves.  
The lower plots in the two panels disclose the color 
variations only over timescales of $\gtrsim$0.6 kiloseconds.
              }
         \label{litecol}
   \end{figure}

The light curves and hard/soft color ratios for OBS2 and OBS4 are shown
as examples of all four observations in Figure\,\ref{litecol}. 
% Fig 3 introduced here
Note that the light curve exhibits the usual pattern of 
almost uninterrupted flaring. 

  As with previous light curve studies, we find also that 
soft- and hard-energy light curves constructed track one another very
well, giving only occasional significant departures from a mean color ratio.
There is in general a remarkable absence of color hardening during strong
flares or brief near cessations in emission. In fact, most of the largest
color fluctuations occur when no strong flux excursions (flares or 
flux lulls) occur. 
Yet we do find occasional departures from a uniform color.
For example, we found a few sinuous variations in color curves in each 
of the four observations.  An example is depicted for stretches for 
OBS2 and OBS4 - see Figure\,\ref{litecol}. 
In OBS4 sudden hardenings occur over 0.5-1 ksec and are followed by a 
slow return to the mean color over a few ksec. 
 Sometimes hard flux changes are followed by soft fluxes, but not quite
to the same degree (e.g., the ``decays" in Fig.\,\ref{litecol}b). 
At other times they occur when the hard component is nearly constant or 
undergoes apparently stochastic fluctuations (Fig.\,\ref{litecol}a).  
In almost all these cases the soft component changes more slowly, on 
timescales of 100--1000 seconds, in the fashion shown in these plots.

\begin{figure}
   \centering
   \includegraphics[width=6cm,angle=90]{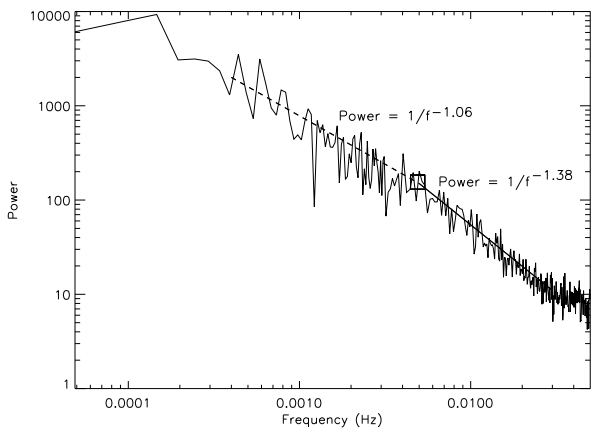}
      \caption{
      Sum of power spectra of our four {\it /XMM/EPIC pn} observations over 
0.8-10\,keV of $\gamma$\,Cas. 
The binning in time over 10 seconds prevents the spectrum from 
being followed out to its white noise component.  Regression lines are drawn
with slopes noted through two frequency ranges. A break in the slopes is noted
by a square symbol at a frequency of 0.005 Hz.
              }
         \label{pwrsp}
   \end{figure}

The power spectrum,
summed over our four observations, is exhibited in Figure\,\ref{pwrsp}. 
% Figure 4 introduced
It shows the beginning of a logarithmic slope just exceeding -1 for $\ge$ 30s 
flare-like excursions. 
At about 0.005 Hz it breaks to a shallower slope. This is consistent 
with the break at 1-3$\times$10$^{-3}$ Hz reported by Robinson \& Smith (2000). 
We characterize the occasional variations on timescales longer than flares
as ``undulations." SRH and RS have shown that these features are most
likely due to the passage of translucent clouds 
forced into corotation over the Be star. They suggested that these 
are responsible for the continued rise in power above the break in the
slope of the power spectrum. Looking to shorter timescales, we performed
an autocorrelation analysis on the integrated light curves as well as
light curves separated into binned soft and hard fluxes. However, we 
found that none of them exhibited quasi-periods. Also, we found that
to a precision of 1--2 sec we could find no evidence of time delays
between the hard and soft flux time series.

\subsection{Spectral analysis: primary parameters}
\label{xrspct}

\noindent {\it Description of XMM-Newton spectra}

  The 2010 {\it XMM-Newton} spectra appear similar to
those obtained by Chandra in 2001 (Smith et al. 2004) and the {\it XMM-Newton}
in 2004 (Lopes de Oliveira et al. 2010, ``L10"). The spectrum is ``hard," 
i.e., most continuum flux is emitted at energies of a few \AA ngstroms
or lower.  The long wavelength continuum 
is attenuated much like the Chandra spectrum and in 
contrast to the unattenuated 2004 {\it XMM-Newton} spectrum. 
The visible emission lines in the long wavelengths of the {\it RGS} 
spectral detectors are the Ly\,$\alpha$-like lines of even-nucleons 
Mg\,XII, Ne\,X, and O\,VIII, and the $i$ and $f$ components of the so-called
$fir$ complex of the ground-level helium-like ions Ne\,IX and O\,VII. 
The Ly$\alpha$-like emission lines of 
the Fe K shell ions Fe\,XXVI and Fe\,XXV are likewise visible in the short 
wavelength spectra of the {\it EPIC}/$MOS$ and $pn$ detectors.  
An additional emission feature at $\approx$1.94\,\AA~ arises from 
atoms of less excited Fe ions, the so-called Fe K fluorescence feature.
Similarly, in three of the four {\it RGS} camera spectra
we found faint corresponding features due to Si\,K ion fluorescence,
as previously reported by Smith et al. (2004, ``S04"). In addition,
the {\it RGS} spectra exhibit a number of emission lines of Fe\,XVII 
and Fe\,XVIII (Fe L shell) ions. The quantitative analyses of our line 
strengths and broadenings were performed independently for spectra recorded 
by the {\it pn} and {\it MOS} detectors at short wavelengths 
and for the $RGS1$ and $RGS2$ detectors at long wavelengths because the 
instrumental responses and calibrations for these cameras are different. 
Following {\it XMM-Newton} mission recommendations we have performed
quantitative analyses with {\it XSPEC} using all individual {\it EPIC} and 
{\it RGS} spectra simultaneously.

   \begin{figure*}
   \centering
   \includegraphics[bb=2cm 1.3cm 10.1cm 27cm,clip=true,width=5cm,angle=-90]{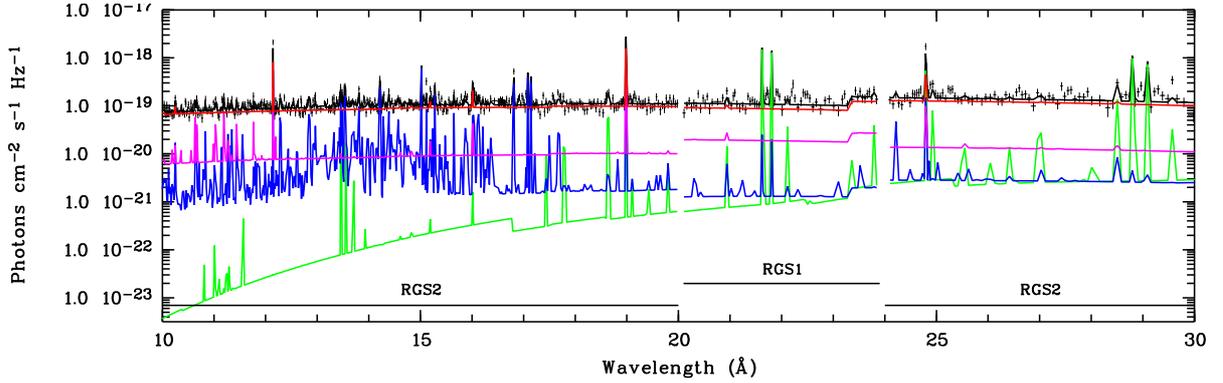}
      \caption{
      Color-coded ``unfolded" model spectra in the range 10--30\,\AA~ 
computed for RGS1 and RGS2 for the fitted model of OBS3. 
The green, blue, purple, and red denote the normalized emission spectra 
of plasma components $k$T$_1$, $k$T$_2$, $k$T$_3$, and $k$T$_4$, each with the 
single values given in Table\,4. The black line is the observed spectrum.
The model from the RGS2 detector was used for all wavelengths except for 
the 20-24\,\AA~ gap, and for this gap the RGS1 model was used. 
This accounts for small discontinuities in the emissions at these wavelengths 
for the $k$T$_2$ and $k$T$_3$ plasma components.
The $k$T$_3$ and $k$T$_4$ emission are almost as well described 
by a broad DEM centered at a high energy. 
              }
         \label{rgsunf}
   \end{figure*} 

\vspace*{.15in}

\noindent {\it Using XSPEC to determine multi-temperature models}

 We used the same suite of {\it ftools} and {\it XANADU/XSPEC} tools to fit
our {\it XMM-Newton} spectra that L10 used --  mainly 
{\it mekal,} {\it vmeka,} and {\it cevmkl.} 
We also used the XSPEC model {\it bvapec} to determine measures of line 
widths and certain elemental abundances. 
All of these models assume one or more optically thin and thermal
plasma components that are also in coronal ionization equilibrium. 
The {\it mekal} model families are then parameterized by a series of 
mono-thermal plasma components and their
normalizations (proportional to Emission Measures, ``EMs").
The normalizations in the various cameras were independent of one 
another and all were kept free during the fit. This procedure was 
adopted to assure that each individual normalization takes into account 
the difference in calibration of the cameras. 
Ultimately we adopted the mean value of the normalization of each 
camera and for each plasma component to compute an emission measure
from its recorded spectrum. However, in doing so we neglected the weak and
uncertain high energy response of the RGS cameras, which can otherwise bias
the result.
A Gaussian emission feature was added to the model to parameterize
the visible Fe fluorescence line at 1.9\,\AA.~ We also included necessary
multiplicative photoelectric absorption terms, parameterized by the term
``$n_{H_i}$," which correspond to the equivalent hydrogen columns. 
These parameters are combined in a simultaneous  
solution to fit the continuum and emission lines. 
In general, a number of iterations are required for convergence.
We experimented with different starting temperatures to avoid converging to
false solutions. 

 As in L10 we tried to fit the $\gamma$\,Cas {\it XMM-Newton} spectrum with a
continuous Differential Emission Measure (DEM) defined as a cooling flow model
(e.g., {\it cevmkl})\footnote{The differential emission measure for cooling
flows is given by dEM = [(T/T$_{\rm max}$)$^{\alpha}$](dT/T$_{\rm max}$). 
For adiabatic flows $\alpha$ $\equiv$ 1.}. While this is sometimes considered 
the preferred X-ray emission description for accreting white dwarfs, 
it does not work for our spectra. For example, for adiabatic 
flows and T$_{\rm max}$ allowed to float to 28 keV, the continuum fit is
poor, giving a reduced $\chi^{2}$ = 1.69. For a model where the $\alpha$ index
is permitted to float to its optimum value of 1.31, the fit is still poor,
$\chi^{2}$ = 1.52.

%%%%%%%
\begin{table*} 
\caption{Best parameters from X-ray Model of the form: $(T_1+T_2+T_3+T_4)*n_{H_a}+(T_4+gauss)*n_{H_b}$.}              
\label{tbl:parameters}        \centering           \begin{tabular}{lcccc}      
\hline                    
\hline\\[-2.2ex]

Parameter & OBS1 & OBS2 & OBS3 & OBS4 \\
\hline
$n_{H_a}$ (10$^{22}$ cm$^{-2}$) & 0.133$_{-0.001}^{+0.001}$ & 0.124$_{-0.001}^{+0.001}$ & 0.090$_{-0.001}^{+0.001}$ & 0.128$_{-0.001}^{+0.001}$\\
$kT_{1}$ (keV)                             & 0.115$_{-0.002}^{+0.002}$  & 0.110$_{-0.002}^{+0.002}$ & 0.104$_{-0.001}^{+0.001}$ & 0.111$_{-0.002}^{+0.002}$\\ 
EM$_1$ (10$^{55}$ cm$^{-3}$)  & 0.14 &  0.10 &  0.21  &   0.12 \\

$kT_{2}$ (keV)                             & 0.60$_{-0.02}^{+0.02}$        & 0.64$_{-0.02}^{+0.02}$      & 0.63$_{-0.01}^{+0.01}$ & 0.62$_{-0.01}^{+0.01}$\\

EM$_2$ (10$^{55}$ cm$^{-3}$) & 0.33  & 0.27 &   0.37  & 0.32 \\

$kT_{3}$ (keV)                             & 3.63$_{-0.19}^{+0.22}$        & 4.30$_{-0.19}^{+0.22}$      & 4.20$_{-0.14}^{+0.14}$ & 3.43$_{-0.13}^{+0.13}$\\

EM$_3$ (10$^{55}$ cm$^{-3}$) &  0.18  &  0.29  &  0.45  &  0.22   \\

$kT_{4}$ (keV)                             & 14.66$_{-0.39}^{+0.31}$      & 14.23$_{-0.22}^{+0.25}$    & 15.71$_{-0.23}^{+0.22}$ & 13.51$_{-0.16}^{+0.18}$\\

EM$_{4_{(n_{H_a})}}$ (10$^{55}$ cm$^{-3}$) &  3.0  &  2.6  & 3.5  &  2.6   \\

EM$_{4_{(n_{H_b})}}$ (10$^{55}$ cm$^{-3}$) &  0.7 &  1.0  & 0.8  &  0.7   \\
$f_T$ (erg\,cm$^{-2}$s$^{-1})$ & 2.63$\times$10$^{-10}$ & 2.49$\times$10$^{-10}$   & 3.89$\times$10$^{-10}$  & 1.79$\times$10$^{-10}$ \\

$n_{H_b}$ (10$^{22}$ cm$^{-2}$) & 36.3$_{-3.3}^{+4.6}$            & 57.0$_{-4.2}^{+3.8}$         & 59.6$_{-2.6}^{+3.5}$  & 73.7$_{-3.0}^{+3.2}$\\

redshift      &         0.0018  &     0.0005 &    0.0007   &     0.0003 \\  
$\chi^2_{\nu}$/d.o.f.                      & 1.15/3189                                     & 1.18/2877                                 & 1.18/3663  & 1.20/3441\\

\hline                                  \end{tabular} 
\begin{list}{}{}
% \item[$^{\mathrm{1}}$]
\item The following elemental abundances were fixed to nonsolar values: $Z_{FeK\alpha}$ = 0.18$\times$$Z_{Fe, \odot}$, $Z_N$ = 2.33$\times$$Z_{N, \odot}$, $Z_{Ne}$ = 1.8$\times$$Z_{Ne, \odot}$.
\item[]Nonzero redshift values are not considered statistically 
significant because they are of the order of known seasonally-induced shifts 
in the instrument (e.g., Pollack 2010).
~Entries marked ``f$_T$" are unabsorbed fluxes in the energy range 0.2-12\,keV.
\end{list}
\end{table*}

%%%%%%%
Next we ran multiple-component optically thin thermal models with
{\it mekal} for the five {\it XMM-Newton} spectroscopic cameras. 
% TABLE 5
The final results are shown in Table\,4. These modeling results were 
similar to the results of S04 and L10 in that they led us 
to an optically thin model with thermal plasma components and
two absorption columns. The solution for the cool
plasma component denoted by $k$T$_2$ is an increase of 0.8 keV from the
L10 model - a statistically significant result due to an actual spectral 
change according to our return to the L10 spectra with our {\it mekal} models. 
The addition of successive temperature components reduced the mean $\chi^2$ 
from 1.70 to 1.48, to 1.28, and to (four components) 1.20.  With each added
component the improvement in fit over certain wavelength regions was marked.
In particular, {\it mekal} models with fewer than four components also 
caused perceptible spurious undulating departures from the observed 
continua that were minimized with the inclusion of the fourth component.
The addition of further components reduced the mean $\chi^2$ by
negligible amounts and were not significant according to statistical
$f$ tests performed in {\it XSPEC.}  
% Therefore, they were not adopted. 
However, like L10 we were
not able to fit both the pn/MOS and {\it RGS} spectra well 
unless we added a second column to attenuate the soft energies. 
We may parameterize our best {\it mekal} model with the equation:

\begin{center}  

\begin{displaymath}
phabs_{a}\ast (mekal_1 + mekal_2 + mekal_3 + mekal_4) + 
\end{displaymath}
\begin{equation}  
~~~~~~ + phabs_{b}\ast(mekal_4 + gaussian)~~~, 
\end{equation}  
\end{center}
\noindent where the {\it mekal} subscripts 1-4 parameterize the temperatures
$k$T$_1$-$k$T$_4 $ of the four thermal plasma components of a 4-T model. 
The subscripts in {\it phabs} express the two independent column 
densities, $n_{H_a}$ and $n_{H_b}$. As we will see, these values can
be very different. We stipulate that it is possible in principle 
that the $k$T$_{2-4}$ components could be represented in 
the $n_{H_b}$ term as well, 
but their fluxes are so much weaker that we cannot solve for 
them as additional independent variables. 
In Figure\,\ref{rgsunf} we exhibit the unfolded model solution, including 
%Figure 4 introduced 
the normalizations, as an example for the range of 10--30\,\AA\ in OBS3. 
This gives a good feel for the relative contributions among the various 
plasma components for both the line and continuum spectra.

  Figure\,\ref{rgsunf} shows that even in the soft X-ray region the
hot kT$_4$ component dominates the flux. In $\S$\ref{xrlite} we noted
that the hard/soft ratios can change over timescales of 100-1000 seconds.
Despite the dominance of the hot emission component in the soft regime,
the cool kT$_1$ plasma can sometimes make its presence 
felt independently of the hot one.  The most 
%likely means for producing soft X-ray variability is the presence of
%cold gas that absorbs soft X-rays along the line of sight toward the hot
likely means for producing soft X-ray variability is for the line of sight
column of matter to change to the hot emission sites behind it.
%emission sites in the background. 
If one assumes that the hot component fluxes decreased by free cooling,
i.e., proportional to the reciprocal of the density, this timescale 
gives an electron density of $\sim$10$^{11}$ cm$^{-3}$. 

 Whereas components $k$T$_2$ and $k$T$_1$ are critical to the fits of lines of
the Fe L-shell, N, and C ions, the $k$T$_3$ component is the least stable in
our solutions. One reason for including $k$T$_3$ is that it easily produces
the Fe\,XXV Ly$\alpha$ line and contributes to part of the continuum. 
However because 
the continuum flux spectrum of our $k$T$_3$ runs parallel to the $k$T$_4$ in 
a logarithmic plot (Fig.\,\ref{rgsunf}), it is possible that this component 
is an artifact of what is actually a continuous Differential Emission 
measure (DEM) parameterized by $k$T$_4$ but extending down to a few keV. 
To test this possibility we ran an {\it XSPEC} add-on package written
by K. Arnaud called {\it vgadem,} which is a DEM response shaped like a
Gaussian, with a peak emissivity at 12.25 keV and a sigma of 5.47\,keV. 
The values of the $k$T$_1$ and $k$T$_2$ were essentially identical with 
those determined with the {\it mekal} model.
The resulting model replaces the monoenergetic emissivity terms, $k$T$_3$ + 
$k$T$_4$, in the parenthesis term of equation (1) and $k$T$_4$ in the second 
term with one having a broad DEM.  For our test case (OBS1) this model
gave an acceptable $\chi^{2}$ of 1.19 for the global fit. 
However, we found this solution to be unstable in convergence, and it 
slightly underpredicts the strength of the Fe\,XXVI line. In all, it
does not perform as well as the {\it mekal} 4-T model over all wavelengths.
We had a similar result running {\it vgadem} with the 2004 spectrum
and abandoned this model. From these trials we could not corroborate 
our hypothesis that the $k$T$_3$ plasma is an unnecessary product of our 
reference model.

We pursued this investigation by comparing our {\it mekal} 4-T and 3-T models 
and excising the wavelengths around the Fe\,XXV line.  We found that the
inclusion of the $k$T$_3$ apart from these wavelengths causes the 4-T model 
to fit the continuum slightly better ($\chi^2$ from 1.19 to 1.15). Thus, the 
$k$T$_3$ in {\it mekal} seems needed for an ideal fit to the continuum too. 
However while introduction of the $k$T$_3$ improves the fit for Fe\,XXV
in our reference model, the fit for Fe\,XXVI 
is actually made slightly worse than for the reference model.
% The correlation of 
% the values of the kT$_3$ values themselves among the 2001, 2004, and 2010
% spectra with the values of k$T_4$ further argues that the kT$_3$ and kT$_4$
% components actually refer to the same plasma.
To conclude, although we cannot provide an objective reason for preferring 
the 3-T {\it vgadem} DEM model over the reference model, the {\it mekal} 
model with $k$T$_3$ $\approx$ 4 keV actually appears not to be ideal 
either. We suspect a flatter DEM with cutoffs at low and high energies
could improve these fitting deficiencies.  In fact, there are physical 
reasons for believing that the high energy source for the X-ray 
emissions of this star is not single-valued in energy even as there
are reasons for suspecting that the ``$k$T$_3$ component" is in fact not 
independent of the $k$T$_4$ one.

  We give the parameters found in the {\it mekal} solutions represented 
by equation (1) in Table\,4. The solution also stipulates the anomalous
abundances found. Some 80\% of the flux is emitted by a hot 
%TABLE 4
component we designate $k$T$_4$.
This percentage increases to about 87\,\% for the case that $k$T$_3$ represents
an extension of $k$T$_4$. 

 The $k$T$_4$ component dominates emission at all wavelengths, so much so
that it serves to weaken color ratio changes represented in
Figure\,\ref{litecol} when 
the $k$T$_4$ component alone does not vary. This emission also dilutes
the equivalent widths (EWs) of some lines produced in cooler plasmas. 
Figure\,\ref{rgsunf} shows the unfolded model RGS spectra for OBS4 from 
the 4-temperature model described in Table 4. We remind the reader that the 
$k$T$_4$ flux is composed of a flaring and a dominant basal component.

\subsection{Analysis of secondary parameters}

\noindent {\it Column densities, abundances, and line broadening:}

We list additional details emerging from our analysis: 
\begin{itemize}
\item the $n_{H_b}$ values are the largest documented to date
and are a few times larger than the 1$\times$10$^{23}$ cm$^{-2}$ found by S04. 
The values change by $\approx$2$\times$ among the four spectra. 

 \item the ratios of
  volumes associated with the ``lightly absorbed" ($n_{H_a}$) and 
  ``heavily absorbed" ($n_{H_b}$) columns (given by the respective EMs
     in Table\,4 and in tables in S04 and L10) remain the same during
  2001, 2004, and 2010. From Table\,4 these ratios are
 25$^{+5}_{-15}$\%, 23${\pm 3}$\%, and 27${\pm 5}$\%, respectively.
%  correspond to 25$^{+5}_{-15}$\%, 23\%, and here 25${\pm 5}$\%, respectively,

\item the {\it bvapec} task was used to determine abundances of 
  elements from all the strong lines. Like several previous studies, the Fe 
  abundance derived from the L shell ions is close to the solar value. 
  The abundance determined from the two Fe K shell ion lines is 
  0.18$\times$ the solar value. This is statistically significantly 
  higher than the even lower value of 0.12 found by L10.  We used
  {\it bvapec} also to determine Ne and N abundances of 1.80 and 2.33
  times solar, respectively, from their Ly\,$\alpha$ line strengths. 
  This is slightly more moderate than the 2.63${\pm 0.78}$$\times$ 
  and 3.96${\pm 0.28}$$\times$ solar values that L10 found.  
  We found no statistically significant changes in these 
  abundances among the four 2010 observations.

\item 
  the {\it XSPEC} model {\it bvapec} was used on the strongest line
  profiles in the RGS spectra to determine Gaussian line broadening. 
  Our solutions were dominated by the strong O\,VIII line. We found FWHM 
  values in the range 540-950 km\,s$^{-1}$. These values are marginally larger
  than the moderate values of 400 km s$^{-1}$ and 478 km s$^{-1}$ found
  by S04 and L10, respectively.

\item Fe\,K and (barely) Si\,K fluorescence features are present. 
   The EW of the Fe feature is -35m\,\AA~ to -50\,m\AA. Although these
   features were present in earlier high resolution spectra, their 
   strengths in the 2010 spectra are the highest reported  so far. 

\end{itemize}

\noindent {\it Spectra derived from low and high flux time intervals:}

  In order to understand how key fitting parameters 
change when $\gamma$\,Cas's emission changes between low to high states
(usually due to flaring), we split 
up the good time intervals according to low and high fluxes (relative to
the means) and reconstructed spectra into low- and high- flux pairs. We
discovered that the low  and high  flux times were strongly grouped together,
creating split segments of times. The duration of these various segments
was from 3--30 minutes. This timescale is longer than durations of flares.

The flux differences between these pairs were statistically significant, but 
only for the $k$T$_4$ plasma component. Except for OBS1, the high-flux spectrum
was several percent hotter than the low-flux one. The largest difference in
the solution for the $k$T$_4$ component 
occurred for OBS3. In this case the $k$T$_4$ values for the high and low
flux values were 16.08 keV and 14.59\,keV, which, given the errors 
of ${\pm 0.59}$\,keV, is statistically significant at more than two sigma.
% **(I do not remember if we have obtained the errors (I think so). Anyway, it is important to show them in order to say that the values are statistically significant.)**
(We repeated this experiment for the 2004 {\it XMM-Newton} 
observation by splitting it
into two equal time segments and examining the low and high flux states in each
 half. We found no evidence for temperature differences for this observation.)
Differences in the column densities between the pairs were not significant. 
 
  Even so, we found several statistically significant differences  among
the high-/low-flux spectra among several emission lines, most especially
those arising from the Fe\,L shell ions, an emission feature at 16\,\AA~
(the wavelength of O\,VIII Ly\,$\beta$) and a neighboring Ne\,X line, and also
a blend of the N\,VII Ly\,$\alpha$ and a nearby N\,VI line at 24.8\,\AA.~
The general pattern is that these lines are stronger in the low-flux spectra.
To shed light on these changes, we constructed unfolded spectra
for each of the 8 high/low-flux spectra  (two for each one of the four 
observations), according to their normalizations.
Examples of this unfolded low-flux spectrum for
OBS2 and OBS3 are shown in Figure\,\ref{fe17n7unf}.  
% Figure 6 introduced 
The prominence of these lines was actually due to a diminished dominance 
of the $k$T$_4$ continuum, but not the lines, for the low spectrum
fluxes. In other words, the spectra of all the cooler components
then ``peek through" the dominant continuum emitted by $k$T$_4$.
The result is a highly dynamically changing spectrum. The details of the
low/high flux spectrum differences can be summarized in the following:

   \begin{figure}
   \centering
   \includegraphics[width=6cm,angle=-90]{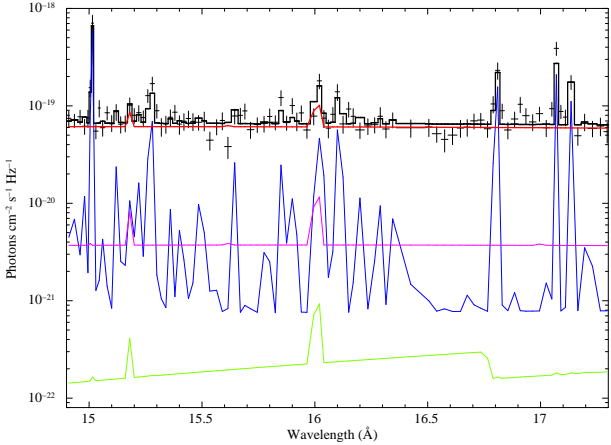}
      \caption{
      Unfolded spectra for the spectral region shown in Fig.\,\ref{FeL},
for OBS3 (same color coding as in Fig.\,\ref{rgsunf}). 
%The example chosen is for our fits to the ``low-flux" spectrum OBS4.  
Most of the lines arise
from the L-shell ions of Fe\,XVII \& XVIII. Note that the strengths of most 
lines depends on the normalization of the kT$_2$ emission component 
(blue) relative to the dominant kT$_4$ component (red).
              }
         \label{fe17n7unf}
   \end{figure}

\begin{itemize}
     \item  differences in strengths of most lines but O\,VIII, e.g.,
     Fe L shell ions of Fe\,XVII and Fe\,XVIII, Ne\,X, 
     and Fe\,XVIII-blended Ly$\beta$/Ly$\alpha$, especially in OBS2 and
     OBS3.  The change is the smallest for O\,VIII Ly$\alpha$ because the
     $k$T$_2$, $k$T$_3$, and $k$T$_4$ plasma components contribute 
     comparable amounts of EW. Therefore, changes among these components' 
     normalizations, for example in the spectra extracted from low and high 
     fluxes, do not affect their equivalent widths significantly.

     \item analysis of unfolded spectra constructed from the high and 
     low fluxes shows why these rapid variations in line strengths occur: 
     the $k$T$_4$ emission component in the low-flux spectrum decreases and 
     allows emission from primarily the $k$T$_2$ component to be more visible. 
     An example is displayed in Figure\,\ref{FeL}b and modeled in 
% Figure 7 introduced here
     Fig.\,\ref{fe17n7unf}. 

   \begin{figure}
   \centering
   \includegraphics[width=6cm,angle=90]{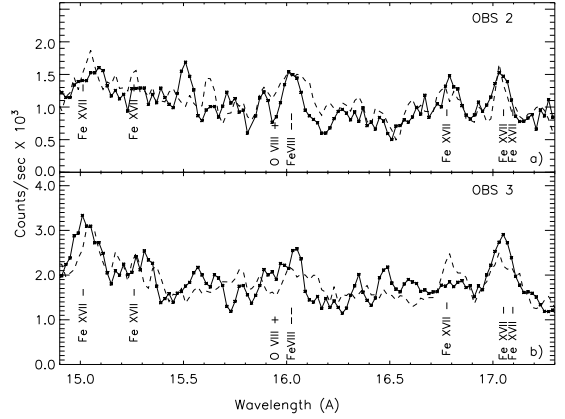}
      \caption{
      {\it XMM-Newton/RGS} spectra 
obtained from high (dashed line) and low (solid) flux good time 
intervals in the 15-17\,\AA~ (Fig.\ref{FeL}) region in which several 
density-sensitive Fe\,XVII lines are present for OBS2 and OBS3. 
The generally greater strengths of these lines 
and the 16.8\,\AA/17.0\,\AA~ ratio are 
discussed in the text. These spectra are smoothed over two 2\,m\AA\ bins.
              }
         \label{FeL}
   \end{figure}

The change in $n_{H_b}$ for low/high (OBS3) is of the same
     order as the deviations around the observational averages. 
\end{itemize}  

\noindent {\it Other lines and density diagnostics:}
    
      Similar to S04 and L10, where the $fir$ 
      (forbidden/intercombinational/radiative) triplet complex is 
      observed, in this case for N\,VI and O\,VII, we find that the ratio
      G $\equiv$ $(i + f)/r$ $\approx$ 1, and this means that the dominant 
      process for formation of these triplet lines is consistent with 
      the dominance of collisions in determining the relative level 
      populations. Likewise, we find the ratio $f/(i+r)$ $\sim$ 0. This might 
      be explained by the metastability of the upper level (1s2s~$^{3}$S$_1$)
      of the forbidden transition. However, it is at least as likely that
      this level is radiatively depopulated by intense UV flux from a 
      nearby hot star, in which case the utility of the forbidden component 
      as a density diagnostic is vitiated. Thus, as with the former studies, 
      we cannot draw conclusions about density from these features.

         Other density-sensitive diagnostics are available among
      several Fe\,XVII lines at wavelengths accessible to both 
      the {\it Chandra/HTEG} and {it XMM-Newton/RGS} wavelength regions. 
      These have been studied empirically and theoretically in detail by 
      Mauche and collaborators (Mauche 
      et al. 2001, 2004) in the well observed Chandra/HTEG spectra of the
      mCV (polar) star AE Aqr.  The lines in this source appear to be emitted 
      in plasma with thermal conditions similar to those of the $k$T$_2$ 
      plasma of $\gamma$\,Cas. The ratios formed by these line strengths 
      are more useful for these purposes than the $fir$ triplets because 
      the atomic level populations are found to be affected by UV 
      photoexcitations by only $\approx$10\% or less (Mauche 2002).

       In the competition between radiative and collisional deexcitations 
       of the upper atomic levels, high densities mainly affect 
       the lines arising from transitions from the lowest member of an 
       excited manifold. For the resonance Fe\,XVII lines we can 
       observe these terms of these manifolds correspond to the 2p$^{5}$3d 
       and 2s$^{2}$3s terms. The observable transitions most affected are 
       15.26\,\AA\ and 17.10\,\AA, respectively.  Both of these are too
       weak to be used in our analysis, so we must resort to alternative
       diagnostics. 
       
       Fig.\,\ref{FeL} exhibits the principal Fe\,XVII ion lines visible in
       the 15-17 \AA~ region. The spectra shown are averages of features
       recorded in both RGS cameras and smoothed over two wavelength bins. 
       The line ratio formed by 17.10\,\AA~ (magnetic quadrupolar 
       transition) and 17.05\,\AA~ is a density diagnostic
       (Klapisch et al. 1978, Mauche et al.  2001). 
       We estimate that in our spectra the ratio I(17.1\AA)/I(17.05\AA)
       lies in the range of 0.0-0.4 (with a value near zero most likely). 
       This means that
       the electron density is of order 10$^{14}$ cm$^{-3}$ or higher,
       according to Mauche et al. (2001). For
       OBS2 the ratio I(16.8\AA)/I(17.0\AA) $\approx$ 1. From the work of
       Mauche et al. this ratio implies a density near 10$^{13-14}$ cm$^{-3}$ 
       for both the high-flux and low-flux spectra. However,
       for OBS3 this same ratio has values $\approx$1 for the
       high flux spectrum and decidedly less than 1 for the low flux spectrum. 
       The last ratio indicates  N$_e$ $\gtrsim$ 10$^{14}$ cm$^{-3}$.
       (This same condition occurs for both the low and high spectra of
        OBS4, although it is not shown.) 
       This difference suggests that the densities of
       the emitting plasma can sometimes fluctuate even on timescales of 
       several minutes. Moreover, that the low-flux spectra 
       constructed from any of the other observations shows no comparable 
       weakness of 16.78\,\AA\ and that the higher density range is more 
       typical.  Altogether, the Fe\,L ion lines suggest high densities for
       the $k$T$_2$ plasma component. These electron 
       densities are centered near 10$^{14}$ 
       cm$^{-3}$ and can fluctuate around a mean value of 
       10$^{13-14}$ cm$^{-3}$ on either short or long
       timescales. The densities for this plasma component are 
       similar to those found for the high temperature flares from 
       independent plasma cooling arguments (SRC).
       This high density constitutes one of the more 
       surprising results we have found in this study.

\section{Discussion}
\label{discss}

\subsection{\it Connections to the circumstellar environment} 

   Our analysis of CHARA {\it MIRC} and {\it VEGA} data resulted
in similar disk solutions.
There is also an indication that matter
may be dispersed beyond the disk and is visible in the near-IR.
As for the disk itself, its extent is larger in the H-band than the R-band
(near H$\alpha$), which is consistent with the higher disk opacity 
expected in the H-band.

 In addition, only the {\it MIRC} ring model gave acceptable errors.
Certainly, a ring model cannot be ruled out from our study.
However, if future LPOI work demonstrates that the disk reaches the star,
%However, in that event any future result that the disk reaches the  star
this result will have to be reconciled with the presence of spectral 
{\it msf} and brief UV continuum light dips
These signal the presence of corotating structures close to the Be star. 
It would seem that shear forces would 
destroy a high density structure in a transition zone between 
corotating regions and a Keplerian disk extending inward that 
intrudes upon them. Thus, we expect the shears would 
force the inner disk transition to evolve to a ring structure. 
It may be important that Hony et al. (2000) found that the weak 
Brackett emission lines in the $\gamma$\,Cas spectrum are broadened to
$\approx$ 550/sin\,i km\,s$^{-1}$ and gave arguments why the
broadening is most likely kinetic. This velocity
is significantly above the projected stellar rotation rate. 
This inference could be further evidence of corotating matter close to 
the star.
In the event that the disk touches the star,
the corotating clouds would have to lie
elsewhere, e.g., outside the disk plane at intermediate latitudes.

   \begin{figure}
   \centering
   \includegraphics[width=8cm,angle=90]{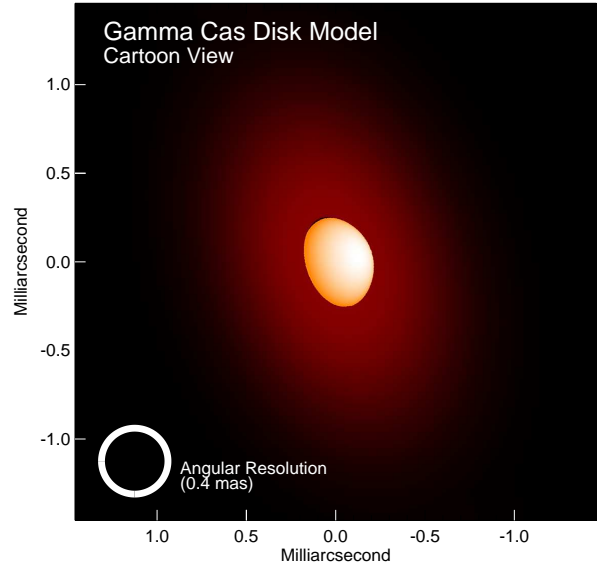}
      \caption{
      Artistic rendition depicting the IR H-band emission 
determined for $\gamma$\,Cas. We show the derived extent, elongation,
and position angle of the Gaussian disk. The inner region is optically thick. 
Neither partial obscuration of the star by the disk nor the star's 
gravity-darkening are included in the MIRC model.
              }
         \label{johnm}
   \end{figure}

  Fig.\,\ref{johnm} gives a feel for the geometry of the system
% Fig 8 introduced here
viewed from our vantage point. Models derived from LBOI 
are sufficient to describe the relative contributions of the  
$\gamma$\,Cas star/Be disk system. They indicate that its disk 
contributes 22\% to the $V$ band brightness and only 3--4\% in $B$, 
of the star-disk system (Stee \& Bittar 2001).  (Note that this 
prediction depends on disk conditions observed in 1993 and on assumptions
about the disk-photosphere transition.)
The $B-V$ color variations in Fig\,\ref{ltcrv}a confirm these predictions.  
The Be disk is the only practical place in the system 
where secular and irregular continuum flux variations can originate 
in visible wavelengths.
Moreover, the season-to-season averages shown in Fig.\,\ref{ltcrv}a have a
linear regression slope $\Delta$$B$/$\Delta$$V$ = 0.66\,${\pm 0.02}$. 
This slope is consistent with color/brightness ratio changes observed 
when disks of classical Be stars evolve in time (e.g., Hirata 1982).

 In $\S$\ref{xrspct} we reported that during the 44 day interval of our 
2010 observations the $n_{H_b}$ column increased by about a factor of two.
This rate of increase defines the column 
at which this range can change with time. Altogether a change of 
300$\times$ has been found between 2004 and 2010.
As already stated, our observations covered the orbital interval between
inferior conjunction (Be star closest to the Sun) and RV maximum.
The orbital phase (0.75) at which {\it Chandra} observed the large 
$n_{H_b}$ in 2001 was almost identical to the phase of our first 
2010 observation (0.74). Although it was still high in 2001, this
column was smaller than in any of the 2010 observations.
As the {\it XMM-Newton} sequence progressed toward quadrature, the
$n_{H_b}$ became higher yet. 
These values are summarized in Table\,5.
%TABLE 5

\begin{table} \caption{Density columns (in 10$^{22}$ cm$^{-2}$), 
Gaussian EW of Fe fluorescence, Abundance Anomalies (solar; Fe refers to K-shell)}              
\label{tbl:prop}        
\centering           
\begin{tabular}{cccc|crr}            
%\begin{tabular}{c|r|c|rrr}            

\hline                    
\hline\\[-2.2ex]
Date   & Binary   &  $n_{H_b}$  & EW(FeK) &
Fe & Ne    & N \\
 & Phase &  & (m\AA) &  &  \\
\hline                    

 2001 & 0.75 &  10      &  -19         &  0.10  &      $\sim$1  &  $\sim$1 \\
 2004 & 0.27 &  0.23  &  -10         &  0.12  &     2.63 &  3.96 \\
 2010 & 0.74-0.96 &  36-74   &  -35 to -50  &  0.18  &     1.80 &  2.33 \\

\hline                                  
\end{tabular} 
\begin{list}{}{}
\item ``Fe" refers to Fe abundance from K-shell ion lines.
\item Errors for [Fe], [N], and [Ne]: ${\pm 0.02}$, 
${\pm 0.75}$, and ${\pm 0.28}$, resp. 
\end{list}
\end{table}
%%%%

\begin{figure}
   \centering
   \includegraphics[width=6cm,angle=90]{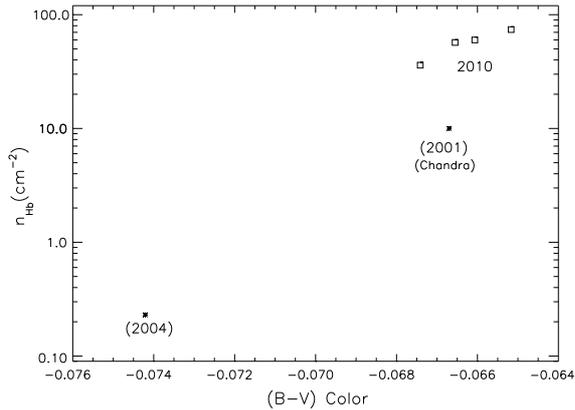}
      \caption{
      The APT-measured $B-V$ colors corresponding to the epochs of the six
n$_{H_b}$ absorption columns determined from soft X-ray flux attenuations. 
The n$_{H_b}$ values increased monotonically during 2010.
The errors in determining n$_{H_b}$ are ${\pm 10\%}$.
              }
         \label{bvnh2}
   \end{figure}

From this it is difficult to maintain that changes seen in $n_{H_b}$ in 2010
can be attributed primarily to binary phase.
If they do correlate with phase it must be in a complicated manner.
Instead, as indicated in Table\,5, $n_{H_b}$ absorption correlates strongly
with the reddening. We exhibit this relation in Figure\,\ref{bvnh2}.
This plot shows that the relation exists not just for the 2001 and 2004 
observations but also for the four in 2010. 
Since we have associated reddening with disk development,
this figure suggests that an association exists between X-ray absorption 
and disk conditions.

  We also find in Table\,5 a suggestion of temporal variance between 
the $n_{H_b}$ column and the strengths of the Fe fluorescence feature, the 
K-shell-derived abundance of Fe, and of [N] and [Ne]. It appears that 
the status of the Be disk is related to several peculiar properties of the 
X-ray emitting plasma.  

   Although it is by no means clear how these physical parameters are
connected, one might begin to connect the dots by noting that the
smaller than unity ratios of $\Delta$$B$/$\Delta$$V$ observed by the
{\it APT} are likely caused by the Be star injecting matter into its
surroundings.  In fact,
variations in $V$ and $B$ of 0.015-0.025 magnitudes over several hours
have been observed on a few occasions (SHV).
% MAY NEED TO CHECK WITH GREG ON HS12 RESULTS
During these events the ratios of the $\Delta$$B$/$\Delta$$V$ were typically
0.85--90, i.e., grayer than longer term variations as recorded in
Fig.\,\ref{ltcrv}a.
Such events suggest a highly dynamic environment in some places
over the Be star, perhaps as associated with failed local ejections
or magnetic prominences.

For more successful ejections to be associated
with the $n_{H_b}$ column, the ejections must: i) occur in lines of sight
subtending a portion of the Be star, ii) be sustained over weeks or months, 
and iii) have a sufficient density that the emission measure (N$_{e}$${^2}$ 
integrated over volume) be visible in continuum light. 
For this study we cannot confirm the validity of these conditions from 
spatial models 
because our LBOI observations are made in optically thick wavelengths 
and thus do not measure changes in volumetric density. 
In consideration of ii) (sustainment), observations of $\gamma$\,Cas in 
the {\it IUE} archives at MAST\footnote{``MAST" represents
the Multi-Mission Archive at the Space Telescope Science Institute (STScI).
%The STScI is operated by the Association of Universities for Research in
%Astronomy, Inc., under NASA contract NAS5-26555. 
Support for MAST
is provided by NASA Office of Space Science via grant NAS5-7584.}
disclose that historically
this star often alternated between prolonged periods of strong and weak states
of the wind emanating from the Be star,
with short transitions of 1-3 months between them.
Generally, though not always, the H$\alpha$ emission EWs were found to 
correlate with the wind state for $\gamma$\,Cas and other Be stars
(e.g., Doazan et al. 1983, 1987). Therefore, it may be that the 
2010 event we have reported for this star is one
such ``wind" episode. We also note that there
is a discrepancy between the range found in the optical depths of
the UV resonance lines and the larger range of n$_{H_b}$ values. 
Lacking a new opportunity for simultaneous UV-X-ray observations of
$\gamma$\,Cas, a resolution of this issue is not at hand.

We make no assumption about what ejection mechanism might be
responsible for the mass loss or tossing episodes.
As suggested in $\S$\ref{dscrp}, a wind driven by radiative and centrifugal
forces alone may play a role in this, but if so it is likely to be modified
by additional processes that are not identified among classical Be stars.
One attribute in the $\gamma$\,Cas wind not found in others 
is a ``flickering" in the violet wing of the Si\,IV and C\,IV resonance
lines on a timescale of 7${\frac 12}$ hours (Cranmer, Smith, \& Robinson 
2000; ``CSR").  This timescale is also seen in the X-ray flux of this star. 
CSR suggested that the rotation into view of X-ray active centers on the
Be star increases the local wind ionization, thereby causing optical
depth changes but not otherwise influencing the wind structure or mass loss
rate.  Whatever the cause, disk diagnostics would be influenced by changes
in the geometry and instantaneous rates of the outflows.  Even so, these
might not change the strength of the X-ray Fe-K fluorescence feature (formed
by a wide range of Fe ions) and certainly not the Fe, N, and Ne 
abundance anomalies.  Although none of this is understood, 
it is at least clear now that these abundances can vary on timescales
of 1--2 years in the X-ray emitting plasma.
This result cannot be understood as a result of recent 
nucleosynthetic processing in the stellar interiors.

  All three high resolution X-ray studies of this star have found it 
necessary to include two absorption columns into the X-ray fitting models.
This raises the question of why it is that no matter what the $n_{H_b}$ 
column density happens to be at any one time, it includes only the same 
fraction, 23-27\%, of the X-ray emitting volumes.  We can address this 
question by supposing that the X-rays are emitted on or very near the surface 
of the Be star from a confined range of latitudes. One can then imagine
that as the flow escapes and crosses our line of sight it will clear 
%and that two other conditions are met: (1) the star disk complex 
%is viewed from an intermediate inclination , as we now know, 
%LBOI work, and (2) the outflow responsible for the $n_{H_b}$ column is 
%always confined to the same equatorial latitudes and feeds the interior
%regions of the Keplerian disk.  If these conditions are met, 
%one can imagine that as the flow emanates radially in our line of 
%sight it will clear 
all but the ``lower" portion of the stellar disk. 
The lower region of the star covered by this outflow, perhaps 25\%, 
would be the X-ray emission volume suffering the 
$n_{H_b}$ absorption. In this picture the fraction of the disk obscured 
by the column would be some function of the confinement of the outflow in 
latitude, the inclination of the star and Be disk, the oblateness of this 
very rapidly rotating star, and the separation of the inner edge of the 
Be disk from the star. With the right selection of parameters the areal 
coverage faction of the $n_{H_b}$ column can remain approximately constant.  

 Our picture must be confronted with additional optical 
photometry and X-ray spectra. The predictions for $\gamma$\,Cas are clearly 
that sudden brightness increases and $B-V$ reddenings should continue to be 
{\it simultaneously} accompanied by increased amounts of expelled particles.  
A portion of the released matter, perhaps this is an episodic wind, is 
directly observable in the soft X-ray domain by its effect on this flux. 
If verified, this picture promises to be more evidence that some, 
and by implication all, of the hard X-ray flux of $\gamma$\,Cas is created 
close to and in front of the Be star.
\vspace*{-.15in}

\subsection{\it A $k$T$_1$-wind connection?}

   Our analyses of the X-ray spectra have not been able to clearly
identify the location of the formation of the secondary 
($k$$T_1$--$k$T$_3$) plasma components. The Fe\,L-shell ion line ratios
yield high density for the $k$T$_2$ component, implying that it is formed near
the primary $k$T$_4$ emission source. As to the cool $k$T$_1$ component,
 L10 discussed the pro's and con's of whether it is associated with the
standard wind of single hot stars. Their conclusion was the
EM1 of this component is larger than found for standard winds but
its temperature is consistent with them.
In this work we can take this conclusion one step further
by pointing out that at least some of the slow I$_{2-10keV}$/I$_{0.3-2keV}$ 
X-ray color variations shown in Fig.\,\ref{litecol} arises from the soft
energy band for which $k$T$_1$ is an important contributor. Because rapid
fluctuations have not been reported from X-rays originating from shocks
in B star winds, we believe this component does not arise primarily
from winds: $k$T$_1$ is somehow part and parcel of the X-ray generation 
processes in $\gamma$\,Cas.

\section{Conclusions}

This study combines contemporaneous information obtained by a 
multidisclipinary campaign on a complicated $\gamma$\,Cas system (Be star 
+ binary orbit + Be disk). We have refined its orbital ephemeris: it 
is in a nearly circular orbit with P$_{orb}$ $\approx$ 203.53 days.  
We also confirmed the Meilland et al. (2007) result using new LBOI
observations in H$\alpha$ light that the 
H$\alpha$-emitting disk rotates at a rate consistent with Keplerian rotation. 
Its extent has been traced out to about one half the Be star's Roche lobe 
radius. We find from optical spectroscopy and imaging that there is no 
longer evidence of a visible one-armed disk structure. 
Moreover, the $V$ an $R$ emission components in the H$\alpha$ line are weaker 
than before the year 2000. These changes are not atypical in the context of 
the behavior of Be disks.

  Yet in other respects $\gamma$\,Cas's recent history is not typical. 
The APT record shows 
an abnormal rate of brightening and reddening of $\gamma$\,Cas, and there 
can be little doubt that this is caused by the interaction of newly ejected 
particles with the disk.  Such events are often observed in 
monitoring Be stars, so this can have little to do with the star's
binarity or the fact that it is emits copious X-rays. Fortuitously,
the activity the APT recorded also coincided with the onset of
our four {\it XMM} observations. As stated in $\S$\ref{xrspct}
the attenuation of soft X-ray flux is caused by an increase in a column
density n$_{H_b}$ that covers ${\frac 14}$ of X-ray emitting 
sites. The rate of increase in n$_{H_b}$ during the 2010
observations is consistent with the changes in n$_{H_b}$ measured in 2001 
and 2004. These column densities 
in turn are correlated with the changes in the $V$ magnitude and $B-V$ 
color but are not related in a simple way to orbital phase.  
We argued that the logical place for the n$_{H_b}$ 
to originate is from the time-dependent efflux from the Be star. 
We noted that these variations occurred without
the additional variable column provided by a 1-armed disk structure.
X-ray studies during the precession of a future V/R cycle could test
conclusively whether the  n$_{H_b}$ variations are associated with
the central Be star.

   Certain other aspects of the X-ray emitting plasma discussed in L10 are
also clearer than before. The multi-thermal component nature of the sites 
seems relatively stable.  We stress that the soft X-ray flux from the 
hot $k$T$_4$ component dominates the contributions
of the cooler components. Even so, one can still see
that the electron densities of the warm plasma component are high, probably
10$^{13}$ cm$^{-3}$ or more. The cool X-ray plasma component clearly forms in 
different volumes but seems to have a higher emission measure, even in 2004,
than is expected for a typical wind of an early B star. In addition, we
noted that the soft X-ray flux sometimes exhibits rapid aperiodic variations.
This is not a known characteristic of Be star winds.

  From the correlation between color, and X-ray column density 
shown in Fig.\,\ref{bvnh2}, it is reasonable
to conclude that ${\frac 14}$ (all?) of the hot X-ray sites lie 
within a high density 
($\gtrsim$10$^{14}$ cm$^{-3}$) medium like the photosphere of a star 
%(probably the Be star} 
and the observer. 
In principle, an alternative interpretation is that the figure 
represents statistical flukes. A second alternative is that the column density
represented by n$_{H_b}$ is associated with accretion onto a dense column 
above a degenerate companion. There are difficulties associated 
with this picture, one of which is the faithful response of
particle densities in the downstream column to an input variation far 
upstream in the Be wind. The picture would assume that the variations in 
the wind, the dense inner regions of the Be disk where the $V$ flux is 
formed, and infall to a secondary some distance away, are all synchronized, 
with at most a short delay. Even assuming this chain of events offers no 
obstacles, such a picture would have to confront the correlations of rapid
variations between X-ray and UV continuum and line fluxes expected to be
visible only in the vicinity of the Be star.

 Aside from its correct interpretation, the reddening-column correlation 
promises for the first time to predict readily observable changes
in X-ray properties of $\gamma$\,Cas.
It will be interesting to see with a future instrument if this relation 
extends to the column densities, abundances, and fluorescence strengths
found for other $\gamma$\,Cas members.

\begin{acknowledgements}

  The quality of this paper was greatly improved by discussions with 
Dr. Maurice Leutenegger. 
We want to also express our appreciation
to Drs. J. Nemravov\'a and P. Harmanec and collaborators for sending us
their manuscript on RVs in $\gamma$\,Cas in advance of publication. 
Dr. Harmanec's insights improved the quality of the paper. 
Dr. J. Nemravov\'a's updates on the status of
Ondrejov H$\beta$ observations filled in an important historical gap.
This work has made use of the BeSS database, operated at LESIA,
Observatoire de Meudon, France: http://basebe.obspm.fr.
We also wish to thank the valuable comments by the referee, especially 
concerning additional interpretations of the wind-disk interaction.
This work was supported in part by NASA Grant NNX11AF71G.
RLO acknowledges financial support from the Universidade Federal de Sergipe
through the MAGIS Program, and also from the Brazilian agency FAPESP 
(Funda\c c\~ao de Amparo \`a Pesquisa do Estado de S\~ao Paulo) 
through a Young Investigator Program (\#2009/06295-7 \& \#2010/08341-3). 
The latter allowed the development his work on this study
during his stay in the Instituto de F\'isica de S\~ao Carlos
of the Universidade de S\~ao Paulo and visit to the United States. 
GWH acknowledges support from NASA, NSF, Tennessee State University, and 
the State of Tennessee through its Centers of Excellence program.

\end{acknowledgements}


\begin{thebibliography}{}

\bibitem[]{Balona09}Balona, L. A. 2009, in Stellar Pulsation: Challenges for
Theory \& Observation, AIP Conf. Ser., 1170, 339

\bibitem[]{Balona11}Balona, L. A., Pigulski, A., et al. 2011, MNRAS, 413, 2403B

\bibitem[]{Berio99}Berio, Ph., Stee, Ph., et al. 1999, A\&A, 354, 203

\bibitem[]{Bjorkman00}Bjorkman, K. S. 2000, priv. commun.

\bibitem[]{Brown08}Brown, J. C., Cassinelli, J. P., \& Maheswaren, M.
2008, ApJ, 688, 1320B

\bibitem[]{Carc11}Carciofi, A. C. 2011, in Active Be Stars, IAUS 272, 325

\bibitem[]{CB12} Carciofi, A. C., Bjorkman, J., et al. 2012, 
ApJ, 744, L15C 

%\bibitem[]{CB06}Carciofi, A. C., \& Bjorkman, K. S. 2006, ApJ, 639, 1081

\bibitem[]{Che10}Che, X., Monnier, J. D., et al. 2010, Proc. SPIE,
7734, 91

\bibitem[]{Clark01}Clark, J. S., Tarasov, A. E., et al. 2001, A\&A, 615, 629 

\bibitem[]{CR89}Collier Cameron, A., \& Robinson, R. D. 1989, MNRAS, 238, 657C

\bibitem[]{Cowley76}Cowley, A. P., Rogers, L., \& Hutchings, J. B. 1976,
ApJ, 88, 911

\bibitem[]{Cranmer00}Cranmer, S. R., Smith, M. A., \& Robinson, R. 2000, ApJ, 
537, 433 (CSR)

\bibitem[]{Delaa11}Delaa, O., Stee, Ph., et al. 2011, A\&A, 529, A87

\bibitem[]{Doazan82}Doazan, V. 1982, in B Stars with and without Emission
Lines, NASA SP-456, 325

\bibitem[]{Doazan83}Doazan, V., Franco, M., et al. 1983, A\&A, 128, 171

\bibitem[]{Doazan87}Doazan, V., Rusconi, L., et al. 1987, A\&A, 182, L25

\bibitem[]{Don99}Donati, J.-F., Collier Cameron, A., et al., MNRAS, 302, 437D

\bibitem[]{Dziem07}Dziembowski, W., Dazyn\'nska, J., \&
Pamyaynykh, A. 2007, Commun. Asteroseismol., 150, 213

\bibitem[]{FM79}Feinstein, A. \& Marraco, H. G. 1979, AJ, 84, 1713

\bibitem[]{Fremat}Fr\'emat, Y., Zorec, J., et al. 2005, A\&A, 440, 305

\bibitem[]{Gies07}Gies, D. R., Bagnuolo, W. G., et al. 2007, ApJ, 654, 527

\bibitem[]{Guedel09} G\"udel, M. \& Naz\'e, Y. 2009, A\&ARv, 17, 309G

\bibitem[]{Harmanec02}Harmanec, P. 2002, Exotic Stars as Challenges to
Evolution, ed. C. A. Tout \& W. Van Hamme, 279, 221

\bibitem[]{Harmanec00}Harmanec, P., Habuda, P., et al. 2001, A\&A, 364, L85H

\bibitem[]{Henrichs83}Henrichs, H. F., Hammerschlag-Hensberge, G, \& Howarth, I. D.
             1983, ApJ, 268, 807

%\bibitem[]{Henry95a}Henry, G. W. 1995a, in ASPC Ser. 79, Robotic Telescopes,
%             ed. G. W Henry \& J. A. Eaton (San Francisco: ASP), 37

%\bibitem[]{Henry95b}Henry, G. W. 1995b, in ASPC Ser. 79, Robotic Telescopes,
%             ed. G. W Henry \& J. A. Eaton (San Francisco: ASP), 44

% \bibitem[]{HS}Henry, G. W., \& Smith, M. A. 2012, in prep. (HS)

\bibitem[]{Hirata82}Hirata, R. 1982, in Be stars, IAU Symp. No. 89 (Dordrecht: Reidel Publ.), p. 41

\bibitem[]{Hony00}Hony, S., Waters, L. B., et al. 2000, A\&A, 355, 187

\bibitem[]{Hony94}Horaguchi, T., Kogure, T. et al. 1994, PASJ, 46, 9

\bibitem[]{Hummel95}Hummel, W. 2000, in The Be Phenomenon in Early Type Stars, 
ed. M. A. Smith, H. F. Henrichs, \& J. Fabregat, ASP Conf Series. 214, 396

% \bibitem[]{Hummel95}Hummel, W. \& Vrancken, M. 1995, A\&A, 302, 751 

\bibitem[]{Itoh06}Itoh, K., Okada, S., et al. 2006, ApJ, 639, 397

\bibitem[]{King87}King, A. R., \& Watson, M. G. 1987, MNRAS, 227, 205

\bibitem[]{LO06} Lopes de Oliveira, R., Motch, C., et al. 2006, A\&A, 
454, 265 

% \bibitem[]{LO07}Lopes de Oliveira, R., Motch, C., et al. 2007, A\&A, 474, 983 

\bibitem[]{LO10}Lopes de Oliveira, R., Smith, M. A., \& Motch, C. 2010, A\&A, 
512, A22 (L10)

\bibitem[]{LO11}Lopes de Oliveira, R. \& Motch, C.  2011, ApJ, 731, 6L 

\bibitem[]{Lucy05}Lucy, L. B. 2005, A\&A, 439, 663L

\bibitem[]{Marco07}Marco, A., Neguerela, I., \& Motch, C. 2007, ASP Conf Ser., ed. S.
 Stefl et al., 361, 388

\bibitem[]{Marco09}Marco, A., Motch, C., et al. 2009, AdSpR, 44, 348 

\bibitem[]{Mauche02}Mauche, C. W. 2002, The Physics of Cataclysmic Variables \& 
             Related objects, ed. B. T.  G\"nsicke et al., ASP Conf. Ser. 261,
             113M

\bibitem[]{Mauche09}Mauche, C. W. 2009, ApJ, 706, 130

\bibitem[]{Mauche01}Mauche, C. W., \& Liedahl, D. A. 2001, ApJ, 560, 992

\bibitem[]{Mauche04}Mauche, C. W., \& Liedahl, D. A. 2004, UCRL-PROC-208729,

\bibitem[]{Meilland07}Meilland, A., Millour, F., et al. 2007, A\&A, 464, 73M

\bibitem[]{Meill07}Meilland, A., Stee, Ph., et al. 2007, AA, 464, 59M 

\bibitem[]{Miroshnichenko02}Miroshnichenko, A. S., Bjorkman, K. S., \& Krugov, V. D.
2002, PASP, 114, 1226

\bibitem[]{Monnier08a}Monnier, J. D., Berger, J., et al. 2008, Proc. SPIE, 5491,
1370

\bibitem[]{Monnier08b}Monnier, J. D., Zhao, M., et al. 2008, Proc. SPIE, 7013, 1

\bibitem[]{Morbey74}Morbey, C. L. \& Brosterhus, E. B. 1974, PASP, 86, 455

% \bibitem[]{MO07}Motch, C., Lopes de Oliveira, R., et al. 2007, in Active OB Stars:
Laboratories for Stellar \& Circumstellar Physics, ASP Conf. Ser., 361, 117

% \bibitem[]{Mourard11}Mourard, D., Berio, Ph., et al. 2011, A\&A, 531, A110

\bibitem[]{Mourard89}Mourard, D., Bosc, I., et al. 1989, Nature, 342, 520

\bibitem[]{Mourard09}Mourard, D., Clausse, J. M., et al. 2009, A\&A, 508, 1073M

\bibitem[]{Mukai03}Mukai, K., Kinkhabwala, A., et al. 2003, ApJ, 586, L77

\bibitem[]{Murakami86}Murakami, T., Koyama, K., et al. 1986, ApJ, 310, L31

\bibitem[]{N11}Neiner, C., de Batz, B., et al. 2011, AJ, 142, 149N

\bibitem[]{Nem}Nemravov\'a, J. 2011, priv. commun.

\bibitem[]{N}Nemravov\'a, J., Harmanec, P., et al. 2012, A\&A, 537, A39

\bibitem[]{OH82}Ochsenbein, F. \& Halwachs, J. L. 1982, A\&AS, 47, 523

\bibitem[]{Okazaki91}Okazaki, A. 1991, PASJ, 43, 75

\bibitem[]{Okazaki01}Okazaki, A., \& Negueruela, I. 2001, AA, 377, 161 

% \bibitem[]{Parmar93}Parmar, A. N., Israel, G. L., et al. 1993, A\&A, 275, 227P

\bibitem[]{Pedretti09}Pedretti, E., Monnier, J., et al. 2009, New A Rev., 53, 353

%\bibitem[]{Perryman97}Perryman, M. 1997, The Hipparcos and Tycho Catalogues, ESA SP-1200

\bibitem[]{Peters82}Peters, G. P. 1982, PASP, 94, 157

\bibitem[]{Peters86}Peters, G. P. 1986, ApJ, 301, L61

\bibitem[]{Pollack10}Pollack, A. 2010, Calibration status of XMM-Newton RGS,\\ 
 http:// xmm.esa.int/external/xmm\_user\_support/usersgroup/20100512/rgs\_calib.pdf

% \bibitem[]{}Pollmann, E. 2007, Be Star Newsletter, 38, 8

% \bibitem[]{Pollmann11}Pollmann, E. 2011, priv. commun.

\bibitem[]{PR03}Porter, J. M., \& Rivinius, T. 2003, PASP, 115, 1153

\bibitem[]{Q97}Quirrenbach, A., Bjorkman, K. S., et al. 1997, ApJ, 479, 477

\bibitem[]{R06}Rakowski, C. E., et al. 2006, ApJ, 649, L111

\bibitem[]{} Rivinius, Th., Baade, D., \& Stefl, S. 2003, A\&A, 411, 229 

\bibitem[]{R07}Rivinius, Th. 2007, in Active OB stars, ed. S. Stefl, S. P. 
Owocki, \& A. T. Okazaki, 361, 219

% \bibitem[]{} Rivinius, Th., Baade, D., \& Stefl, S. 2003, A\&A, 411, 229 

\bibitem[]{RS}Robinson, R. D., \& Smith, M. A. 2000, ApJ, 540, 474 

\bibitem[]{RSH}Robinson, R. D., Smith, M. A., \& Henry, G. W. 2002, ApJ, 575, 435 (RSH)

% \bibitem[]{Roche97}Roche, P. R., Larionov, V., et al. 1997, A\&A, 322, 139 % NEW

\bibitem[]{Safi07}Safi-Harb, S.,  Ribo, M., et al. 2007, ApJ, 659, 407

\bibitem[]{Shafter86}Shafter, A. W., Szkody, P, \& Thorstenson 1986, ApJ, 308, 765S

% \bibitem[]{SS73} Shakura, N. I. \& Sunyaev, R. A. 1973, A\&A, 24, 337 

%\bibitem[]{Sigut09}Sigut, T. A., McGill, M. M., \& Jones, C. E. 2009, ApJ, 699, 1973

\bibitem[]{Slettebak78}Slettebak, A. \& Snow, T. P. 1978, ApJ, 224, L927

\bibitem[]{Smith86} Smith, M. A. 1986, ApJ, 304, 728 

%\bibitem[]{Smith89}Smith, M. A. 1989, ApJS, 71, 357

\bibitem[]{Smith95}Smith, M. A. 1995, ApJ, 442, 812

\bibitem[]{SB06}Smith, M. A., \& Balona, L. B. 2006, ApJ, 640, 491 (SB06)

\bibitem[]{Smith04}Smith, M. A., Cohen, D. H.,  et al. 2004, ApJ, 600, 972 (S04)

\bibitem[]{Smith06}Smith, M. A., Henry, G. W., \& Vishniac, E. 2006, ApJ, 647, 1375 (SHV)

%\bibitem[]{Smith12}Smith, M. A., Lopes de Oliveira, R., \& Motch, C. 2012, in prep.

\bibitem[]{Smith91}Smith, M. A., Peters, G. J., \& Grady, C. A. 1991, ApJ, 367, 302

\bibitem[]{SR99}Smith, M. A., \& Robinson, R. D. 1999, ApJ, 517, 866

\bibitem[]{SR03}Smith, M. A., \& Robinson, R. D. 2003, Interplay of 
Periodic, Cyclic, \& Stochastic Variability in Selected Areas of the 
H-R Diagram, ed. C. Sterken, ASP Conf. Ser. 292, 263

\bibitem[]{SRC98a}Smith, M. A., Robinson, R. D., \& Corbet, R. H. D. 1998a, ApJ, 503, 577 (SRC)

\bibitem[]{SRH98b}Smith, M. A., Robinson, R. D., \& Hatzes, A. P. 1998b, ApJ, 
507, 945 (SRH)

%\bibitem[]{}Stee, P. 2001, priv. commun. % email of Nov 11, 2001; see SB2001, Table4 %   this is also quoted in the RSH02 paper

\bibitem[]{SB01}Stee, P., \& Bittar, J. 2001, A\&A 367, 532 (SB)

\bibitem[]{Stee12}Stee, P., Mourard, D., et al. 2012, A\&A, submitted (Paper 2)

\bibitem[]{Stee98}Stee, P., Vakili, F., et al. 1998, A\&A 332, 268 

% \bibitem[]{SS06}Sturner, S., \& Shrader, C. 2006, AAS HEAD, 9.0130S

\bibitem[]{SS07}Sturner, S., \& Shrader, C. 2007, priv. commun.

\bibitem[]{Tallon08}Tallon-Bosc, I., Tallon, M. et al. 2008, Proc. SPIE, 7013, 44

\bibitem[]{Tanaka}Tanaka, K., Sadakane, K., et al. 2007 PASJ, 59L, 35T 

\bibitem[]{TK94}Telting, J., \& Kaper, L. 1994, A\&A, 284, 515

%\bibitem[]{ten05}ten Brummelaar, T. A., McAlister, H. A., et al., 2005, ApJ, 628,
453

% \bibitem[]{TGV86}Thom, C., Granes, P., \& Vakili, F. 1986, A\&A, 165, L13

%\bibitem[]{Touhami11}Touhami, Y., Gies, D. R., \& Schaefer, G. H. 2011, ApJ, 729, 17

\bibitem[]{Tycner06}Tycner, C., Gilbreath, G.C., et al. 2006, AJ, 131, 2710

\bibitem[]{vL07}van Leeuwen, F. 2007, AA, 474, 653

\bibitem[]{Walk05}Walker, G. A. H., Kuschnig, R., et al. 2005, 2005, ApJ, 
635, L77

\bibitem[]{Welsh93}Welsh, W., Horne, K, \& Gormer, R. 1993, ApJL, 410, L39

\bibitem[]{White82}White, N. E., Swank, J. H., et al. 1982, ApJ, 263, 277

\bibitem[]{Wynn97}Wynn, G. A., King, A., R., et al. 1997, MNRAS, 286, 436

\bibitem[]{Yang88}Yang, S., Ninkov, Z., \& Walker, G. A. 1988, PASP, 100, 233

%   \bibitem[1988]{balluch} Balluch, M. 1988, A\&A, 200, 58

\end{thebibliography}
\end{document}